\documentclass[%
 reprint,
 amsmath,amssymb,
 aps,
]{revtex4-2}

\usepackage{graphicx}
\usepackage{dcolumn}
\usepackage{bm}
\usepackage{hyperref}
\usepackage[mathlines]{lineno}
\usepackage{color}

\usepackage[normalem]{ulem}

\newcommand{\mz}	[0]	{\langle z \rangle}
\newcommand{\vz}	[0]	{\langle \Delta z^2 \rangle}

\newcommand{\kt}	[0] {k_B T}

\newcommand{\dlk}	[0] {\Delta {Lk}}
\newcommand{\lk}	[0] {{Lk}}
\newcommand{\lkn}	[0] {{Lk}_0}

\newcommand{\n}[0]{\nonumber\\}

\newcommand{\dnu}[0]{\Delta\nu}
\newcommand{\dphi}[0]{\Delta\phi}
\newcommand{\dpsi}[0]{\Delta\psi}

\newcommand{\mnu}[0]{\langle\nu\rangle}
\newcommand{\vnu}[0]{\langle\Delta\nu^2\rangle}

\newcommand{\vphi}[0]{\langle\Delta\phi^2\rangle}

\newcommand{\ccnuphi}[0]{\langle\Delta\nu\Delta\phi\rangle}

\newcommand{\ffe}[0]{{\cal F}}
\newcommand{\sfe}[0]{{\cal S}}
\newcommand{\pfe}[0]{{\cal P}}

\newcommand{\omn}[0]{\omega_0}

\begin{document}

\title{Equilibrium Fluctuations of DNA Plectonemes}

\author{Enrico Skoruppa}
\author{Enrico Carlon}%
\affiliation{%
 Soft Matter and Biophysics, Department of Physics and Astronomy, KU Leuven
}%


\date{\today}

\begin{abstract}
Plectonemes are intertwined helically looped domains which form when 
a DNA molecule is supercoiled, i.e. over- or under-wounded. They are 
ubiquitous in cellular DNA and their physical properties have attracted 
significant interest both from the experimental and modeling side. 
In this work, we investigate fluctuations of the end-point distance $z$
of supercoiled linear DNA molecules subject to external stretching forces. 
Our analysis is based on a two-phase model, which describes the 
supercoiled DNA as composed of a stretched and of a plectonemic phase. 
Several different mechanisms are found to contribute to extension fluctuations,
characterized by the variance $\vz$. We find the dominant contribution to
$\vz$ to originate from phase-exchange fluctuations, the transient shrinking 
and expansion of plectonemes, which is accompanied by an exchange of 
molecular length between the two phases. 
We perform Monte Carlo simulations of the Twistable Wormlike Chain and analyze
the fluctuation of various quantities, which are found to agree with the two-phase 
model predictions. Furthermore, we show that the extension and its variance at 
high forces are very well captured by the two-phase model, provided that one 
goes beyond quadratic approximations.
\end{abstract}

\maketitle


\section{Introduction}
\label{sec:intro}
\begin{figure}[b]
\centering\includegraphics[width=0.48\textwidth]{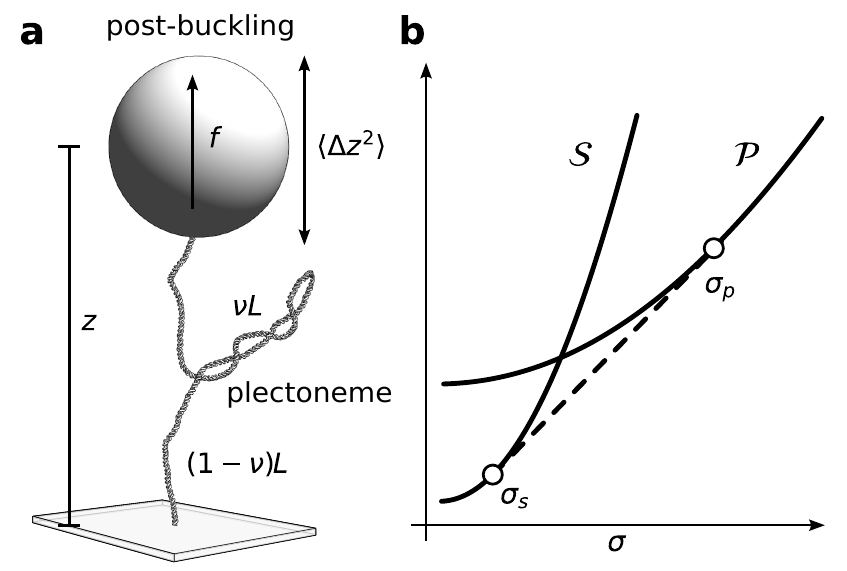}
\caption{(a) Example of plectoneme supercoil formation as induced in MT experiments
by rotation to the end point of a DNA molecule stretched by a linear force $f$.
At post-buckling the molecule phase separates into a stretched and a plectonemic 
phase. The total length of the molecule $L$ is partitioned between the two phases
and $0 \leq \nu \leq 1$ indicates the fraction of length in the stretched phase.
(b) Sketch of free energies per unit length of the stretched and plectonemic 
phases free energies ${\cal S}$ and ${\cal P}$, respectively, as introduced in the 
two-phase model of linear supercoiled DNA buckling \cite{mark95b,mark07}. The dashed line
indicates the double-tangent construction.}
\label{fig:intro} 
\end{figure}

In order to fit into the small volume of a cellular nucleus the $\sim2$m of DNA 
contained in each human cell is arranged in a dense, complex and hierachical manner. 
The bulk of this structural organization is guided by specialized proteins and 
molecular machines. However, on its own DNA 
can achieve a considerable degree of compaction through the formation of plectonemic 
supercoils (see Fig.~\ref{fig:intro}(a)), which are helically intertwined 
regions induced by over- or under-winding the DNA helix (see Fig.~\ref{fig:intro}a).
Besides promoting DNA compaction, plectonemes may induce juxtaposition between 
otherwise distant sites, facilitating the binding of DNA bridging proteins 
\cite{vand19,yan18a,yan18b,yan21}.

Because of its relevance in many biological processes, DNA supercoiling has been 
attracting significant interest for a long time, both from the experimental and modeling side 
\cite{mark95b,mark07,volo79,klen91,volo92,mark94,stri96,fort08,wada09,neuk11,vanl12,ober12,lepa15,fath15,bene15,mate15,iven16,bard18,fosa21,ott20}. 
The properties of DNA supercoils can be probed in vitro by means of single molecule 
experiments such as Magnetic Tweezers (MT) \cite{devl12,lipf11}. 
In these experiments a linear DNA molecule is attached to a solid surface at one side and 
to a paramagnetic bead at the other side (Fig.~\ref{fig:intro}(a)). A magnetic field
is used to apply a stretching force $f$ and to control the total torsional strain of the 
molecule by restricting the rotation of the bead. In a torsionally relaxed DNA, the two 
strands wind around each other $\lkn$ times, corresponding to one turn every $10.5$ base 
pairs, where $\lk$ represents the topological linking number. Rotating the bead away from 
the relaxed state induces an excess linking number $\dlk = \lk -\lkn \neq 0$, which 
can be either positive or negative for over-wound or under-wound DNA, respectively. 
It is convenient to define the supercoiling density $\sigma = \dlk /\lkn$
which is an intensive quantity, independent on the total length of the molecule.
Starting from a torsionally relaxed state ($\sigma=0$) under an applied stretching 
force $f$, one can gradually increase $\sigma$ in a MT experiment. At a certain
threshold $\sigma = \sigma_s$ the molecule buckles - plectonemic supercoils appear and 
the end-to-end distance $z$ drops (Fig.~\ref{fig:intro}(a)). Upon further increase of 
$\sigma$ the plectoneme grows at the cost of the stretched part of the molecule.

Our analysis of linear DNA supercoiling is based on the two-phase model \cite{mark95b,mark07} 
which describes the DNA molecule as being composed of a stretched phase and a plectonemic 
phase, each with distinct free energies, see Fig.~\ref{fig:intro}(b). In this model, 
buckling is analogous to a thermodynamic first order transition. While typical 
theoretical and experimental work on supercoiling in MT focuses on the average signal 
$\mz$, we discuss the fluctuations in the extension $z$ characterized by the variance 
$\vz$. Recent experiments show that a change in $\vz$ can be associated with the formation 
of topological domains induced by proteins that bridge across different sites of 
DNA \cite{vand21}. In addition, insights on equilibrium fluctuations can shed 
light on the rich dynamics of plectonemes \cite{vanl12}.

The aim of this paper is to investigate the properties and origin of fluctuations in 
linear supercoiled DNA. We first discuss the two state model of DNA supercoiling and 
extend it to the analysis of fluctuations. We show that the leading contribution to 
the variance of $z$ are due to {\sl phase-exchange fluctuations}, i.e. the transient 
transfer of contour length between the plectonemic and stretched phase.
Moreover, our analysis highlights that the extension variance $\vz$ is a lot more 
sensitive to the properties of the plectonemic phase than the average extension. 
It is therefore an excellent quantity to be used for the assessment of plectoneme 
free energy models \cite{mark12,eman13}.
The theoretical analysis is supported by numerical results obtained by 
extensive Monte Carlo simulations, which are in general good agreement with
the two-phase model predictions, both for average quantities and their fluctuations.


\section{Fluctuations in extension from the two phase model}
\label{secc:stat_mech}
In order to describe the phenomenology of DNA supercoiling, we follow the commonly 
employed two-phase model-approach that partitions a DNA molecule of length $L$, stretched
by a force $f$ and subject to supercoil density $\sigma$ into two separate phases 
\cite{mark95b,mark07}.
A fraction $\nu$ of the entire molecular length is assumed to be in the stretched phase 
(Fig.~\ref{fig:intro}(a)) maintained at supercoil density $\phi$, while the remainder length 
is in the plectonemic phase with supercoil density $\psi$.
As the total linking number in the system is fixed by the amount of turns
applied to the bead (which means that $\sigma$ is fixed), fluctuations of the 
quantities $\nu$, $\phi$ and $\psi$ 
are constrained to the requirement for the total linking density to equal the sum of the 
contributions within the two domains $\sigma = \nu \phi + (1-\nu) \psi$.
In the following, we will consider $\nu$ and $\phi$ as independent variables,
which fixes $\psi$ to
\begin{equation}
    \psi =  \frac{\sigma-\nu \phi}{1-\nu}.
    \label{eq:psi}
\end{equation}
Following Ref. \cite{mark07} we consider the free energies per unit length of the 
stretched and plectonemic phases ${\cal S}(\phi,f)$ and ${\cal P}(\psi)$, respectively,
such that the combined free energy per unit length becomes the linear combination
\begin{equation}
    \ffe(f,\phi,\nu) = \nu \sfe(f,\phi) + (1-\nu) \pfe(\psi).
    \label{eq:twophase_fe}
\end{equation}
No explicit force dependence is assumed in ${\cal P}(\psi)$ as plectonemes do not 
contribute to the end-to-end extension. While specific choices of these free energies will 
be discussed further on, for the time being we will explore general properties. To reproduce 
the observed phenomenology we only require the free energies to be convex in $\sigma$ and
${\cal S}(\sigma,f) < {\cal P}(\sigma)$ for small $\sigma$, while ${\cal S}(\sigma,f) > 
{\cal P}(\sigma)$ for sufficiently large $\sigma$, reflecting the absence of
plectonemes at small supercoiling densities and their proliferation at
large $\sigma$.  

The partition function of a molecule of total length $L$ subject to a stretching force 
$f$, fixed $\sigma$ and large $L$ is then given by
\begin{equation}
    Z(\sigma,f,L) 
    =
    \int_0^1 d\nu \int_{-\infty}^{+\infty} d\phi
    \,\,
    e^{-\beta L {\cal F}(f,\phi,\nu)}
    \approx
    e^{-\beta L \widetilde{\cal F}(\sigma,f)},
    \label{eq:Z}
\end{equation}
and where the minimal free energy
\begin{equation} 
    \widetilde{\cal F} (\sigma,f) \equiv 
    \min_{\{\phi, \nu\}} \left[ \nu {\cal S}(\phi,f) + (1-\nu)
    {\cal P} \left( \frac{\sigma-\phi \nu}{1-\nu}\right) \right],
    \label{def:calF}
\end{equation} 
is obtained from the condition of vanishing derivatives 
$\partial {\cal F}/\partial{\phi}=0$ and $\partial {\cal F}/\partial{\nu}=0$,
see \cite{mark07}. 
This minimization leads to a double tangent construction (see Fig.~\ref{fig:intro}b)
\begin{equation}
    \widetilde{\cal F} (\sigma,f) = 
    \left\{
    \begin{array}{lcc}
         {\cal S}\left(\sigma,f\right) && 0 \leq \sigma \leq \sigma_s \\
        && \\
        \frac{
        (\sigma_p - \sigma) {\cal S}\left(\sigma_s,f\right) +
        (\sigma - \sigma_s) {\cal P}\left(\sigma_p\right)}
        {\sigma_p -\sigma_s}
        && \sigma_s \leq \sigma \leq \sigma_p \\
        && \\
         {\cal P}\left(\sigma\right) && \sigma \geq \sigma_p. \\
    \end{array}
    \right.
    \label{maxwell_FE}
\end{equation}
In the pre-buckling regime  $0 \leq \sigma \leq \sigma_s$ the minimum corresponds 
to a pure stretched phase, i.e. $\nu=1$ and $\phi=\sigma$. For $\sigma_s \leq \sigma 
\leq \sigma_p$, the free energy is a linear combination of stretched phase and 
plectoneme phase free energies, with a fraction
\begin{equation}
    \label{nu_s}
    \nu_s = \langle \nu \rangle = \frac{\sigma_p - \sigma}{\sigma_p - \sigma_s},
\end{equation}
of molecular contour length contained in the stretched phase. Finally, for 
$\sigma > \sigma_p$, corresponding to $\nu=0$ and $\psi=\sigma$, the plectonemic 
phase fully engulfs the molecule, reducing the end-point extension to zero.
The two-phase model describes buckling as a first order phase transition 
\cite{mark07}. For $\sigma_s \leq \sigma \leq \sigma_p $ the molecule separates 
into stretched and plectonemic domains with {\sl average} supercoil densities 
$\sigma_s$ and $\sigma_p$ in full analogy to a fluid, which phase-separates
into a liquid and a vapor phase with distinct particle densities $n_L$ and $n_V$.
We emphasize that $\sigma_s$ and $\sigma_p$ are average supercoil densities. 
At equilibrium, the supercoil densities ($\phi$ and $\psi$) as well as the 
lengths of the two phases ($\nu$) exhibit fluctuations. It is precisely the 
scope of this paper to analyze the properties of these equilibrium fluctuations.

The experimentally accessible average molecular extension along the direction of 
the force director field is obtained from the total derivative of the free 
energy with respect to the force
\begin{equation}
    \frac{\langle z \rangle}{L} = -\frac{d \widetilde {\cal F}}{d f},
    \label{def:avez}
\end{equation}
while the variance is obtained by the second derivative
\begin{equation}
    \frac{\vz}{L} = -\kt \frac{d^2 \widetilde {\cal F}}{df^2}.    
    \label{def:varz}
\end{equation}


\subsection{Fluctuations in $\phi$, $\psi$ and $\nu$}
\label{sec:theory_gen_tpfluc}

The fluctuations of the supercoiling densities $\phi$ and $\psi$ as well as the fractional 
occupancy of the stretched phase $\nu$ in the post buckling regime can be understood on 
general grounds. Taylor expansion of the free energy per unit length \eqref{eq:twophase_fe}
around the minimum at $\sigma_s$ and $\nu_s$ yields
\begin{align}
    {\cal F} \approx \widetilde{\cal F} 
    + \frac{1}{2} 
    [
    &\ffe_{\nu\nu} (\nu-\nu_s)^2 +
    \ffe_{\phi\phi} (\phi-\sigma_s)^2 
    \n
    &+ 2 \ffe_{\nu\phi} (\nu-\nu_s)(\phi-\sigma_s)
    ],
    \end{align}
where $\widetilde{\cal F}$ is the minimal free energy~\eqref{def:calF}
and $\ffe_{\nu\nu}$, $\ffe_{\phi\phi}$,  $\ffe_{\phi\nu}$ are the second 
derivatives of ${\cal F}$ with respect of $\nu$ and $\phi$.
These form a $2\times 2$ Hessian matrix, which upon inversion gives the 
following results for the fluctuations (for details see 
Appendix~\ref{app:fluctuations})
\begin{align}
    \langle \Delta \nu^2 \rangle 
    &= 
    \frac{\kt}{L(\sigma_p-\sigma_s)^2} \left( \frac{1-\nu_s}{\pfe_{\psi\psi} } +
    \frac{\nu_s}{\sfe_{\phi\phi} } \right),
    \label{eq:gen_vnu}
    \\
    \langle \Delta \phi^2 \rangle 
    &= 
    \frac{\kt}{L\nu_s\sfe_{\phi\phi}},
    \label{eq:gen_vphi}
    \\
    \langle \Delta \nu \Delta \phi \rangle 
    &= 
    \frac{\kt}{L(\sigma_p-\sigma_s)\sfe_{\phi\phi}},
    \label{eq:gen_dnudphi}
\end{align}
where $\Delta \nu \equiv \nu-\nu_s$, $\Delta \phi \equiv \phi- \sigma_s$ and 
where $\sfe_{\phi\phi}$ and $\pfe_{\psi\psi}$ are the second derivatives of the
free energies of the stretched and plectonemic phases with respect to the 
corresponding supercoiling densities. Evaluation at $\phi=\sigma_s$ and 
$\psi=\sigma_p$ is implied. Analogously, the plectoneme supercoiling density 
fluctuates as
\begin{eqnarray}
    \langle \Delta \psi^2 \rangle     = 
    \frac{\kt}{L(1-\nu_s)\pfe_{\psi\psi}}.
    \label{eq:gen_vpsi}
\end{eqnarray}
The variances of the supercoiling densities $\phi$ and $\psi$ are inversely 
proportional to the curvature of the respective free energy at the minimum,
$\sfe_{\phi\phi}$ and $\pfe_{\psi\psi}$, and the average lengths of the two phases, 
$\nu_s L$ and $(1-\nu_s)L$. As in the post-buckling regime $\nu_s$ decreases with 
increasing $\sigma$ (Eq.~\eqref{nu_s}), the variance of the stretched phase supercoil 
density $\langle \Delta \phi^2 \rangle$ increases with $\sigma$. 
Conversely, $\langle \Delta \psi^2 \rangle$ decreases with increasing $\sigma$, 
while the cross-correlator $\langle \Delta \nu \Delta \phi \rangle$ is independent 
on $\sigma$. 

In the pre-buckling regime $0 \leq \sigma < \sigma_s$ the system is in the pure 
stretched phase $\nu_s=1$, such that there are no phase-exchange fluctuations 
$\langle \Delta \nu^2\rangle =0$. When advancing into the post-buckling regime, 
the system traverses the buckling point where the variance in $\nu$ exhibits a 
discrete jump into non-zero fluctuations
\begin{eqnarray}
    \lim_{\sigma\to\sigma_s^{+}}\langle \Delta \nu^2 \rangle - 
    \lim_{\sigma\to\sigma_s^{-}}\langle \Delta \nu^2 \rangle = 
    \frac{\kt}{L}    \frac{1}{\sfe_{\phi\phi} (\sigma_p-\sigma_s)^2}.
    \label{eq:gen_vnu_jump}
\end{eqnarray}
Similarly, fluctuations in $\phi$ and $\psi$ exhibit a discontinuity at the 
buckling point, since $\phi=\sigma$ and $\psi=0$ at pre-buckling.
From Eqs.~\eqref{eq:gen_vnu} and \eqref{nu_s} we find
\begin{eqnarray}
    \frac{d\langle \Delta \nu^2 \rangle}{d\sigma}
    = 
    \frac{\kt}{L(\sigma_p-\sigma_s)^3}
    \frac{\sfe_{\phi\phi}-\pfe_{\psi\psi}}
    {\sfe_{\phi\phi}\pfe_{\psi\psi} },
    \label{eq:gen_vnu_slope}
\end{eqnarray}
which shows that fluctuations in $\nu$ increase throughout the post-buckling regime 
if $\sfe_{\phi\phi} > \pfe_{\psi\psi}$ at the minimum $\phi=\sigma_s$, $\psi=\sigma_p$. 
This is the typical behavior of DNA, as the stretched phase is torsionally stiffer 
than the plectonemic phase \cite{mark07}.

\subsection{Quadratic free energies}
\label{sec:theory_quadfe}

As a concrete and analytically tractable example we consider quadratic free energies 
for the two phases \cite{mark07}
\begin{eqnarray}
    {\cal S}(\phi,f) 
    &=& 
    -g(f) + a(f) \phi^2,
    \label{defS}
    \\
    {\cal P}(\psi) 
    &=& 
    b \psi^2.
    \label{defP}
\end{eqnarray}
While the free energy of the stretched phase of type \eqref{defS} can be derived 
from the Twistable Wormlike Chain (TWLC) \cite{moro97}, the form of the plectonemic
free energy \eqref{defP} is purely phenomenological. It assumes symmetry 
in $\pm\psi$, valid at low forces ($f < 1$~pN) where the DNA has an almost 
symmetric behavior upon over- and under-winding. Eq.~\eqref{defP} can be viewed
as the lowest order expansion of a generic $\pfe(\psi)$. As such, this form will 
be valid for sufficiently small supercoiling densities. For the TWLC 
the coefficients $g$ and $a$ describing the free energy of the stretched phase 
are force-dependent \cite{mark95,moro97,moro98}. Since the system is assumed to be 
fully in the stretched phase at low supercoiling densities, we require $g>0$ 
to ensure that ${\cal S}(0,f) < {\cal P}(0)$. Moreover, to have a transition from 
stretched to plectonemic phase, the free energy of the former has to exceed that 
of the latter (${\cal P} < {\cal S}$) for sufficiently large supercoiling density, 
which requires $a(f) > b$ (this corresponds to $\sfe_{\phi\phi} > \pfe_{\psi\psi}$, i.e. 
an increase in variance with increasing $\sigma$, as per Eq.~\eqref{eq:gen_vnu_slope}). 

For the free energies \eqref{defS} and \eqref{defP} the double tangent construction 
yields the average supercoiling densities of stretched and plectonemic phases \cite{mark07}
\begin{eqnarray}
    \langle \phi \rangle = \sigma_s = \sqrt{\frac{b g}{a(a-b)}} ,
    &\,\,&
    \langle \psi \rangle =
    \sigma_p = \sqrt{\frac{a g}{b(a-b)}}.
    \label{eq:sigsp_quad}
\end{eqnarray}
The free energy in the post-buckling regime ($\sigma_s \leq \sigma \leq \sigma_p$) 
assumes the form
\begin{equation} 
    \widetilde {\cal F}(\sigma,f) = -\frac{ga}{a-b} + 
    2 \sigma \sqrt\frac{gab}{a-b},
    \label{eq:FEquad}
\end{equation}
which is linear in $\sigma$ and depends on $f$ through the force-dependence 
of $g$ and $a$. 

As $\sfe_{\phi\phi}=2a$ and $\pfe_{\psi\psi}=2b$, the fluctuations of Eqs.~\eqref{eq:gen_vnu}-\eqref{eq:gen_vpsi} take the form
\begin{eqnarray}
    \langle \Delta \nu^2 \rangle 
    &=&
    \frac{k_B T \sigma}{2Lg^{3/2}} \, \sqrt{\frac{ab}{a-b}},
    \label{varnu} 
    \\
    \langle \Delta \phi^2 \rangle 
    &=& 
    \frac{k_BT}{2aL \nu_s},
    \label{varphi} 
    \\
    \langle \Delta \psi^2 \rangle 
    &=&
    \frac{k_BT}{2bL (1-\nu_s)},
    \label{corr_dpsi} 
    \\
    \langle \Delta \phi \Delta \nu \rangle 
    &=& 
    \frac{k_BT}{2L} \sqrt{\frac{ab}{g(a-b)}},
    \label{varnuphi}
\end{eqnarray}
where we used \eqref{eq:sigsp_quad}. Note that $\langle \Delta \nu^2 \rangle$ increases 
with $\sigma$, as deduced from \eqref{eq:gen_vnu_slope}: the stretched phase is torsionally
stiffer than the plectonemic phase ($a > b$).

\begin{figure}[t]
    \centering\includegraphics[width=1\columnwidth]{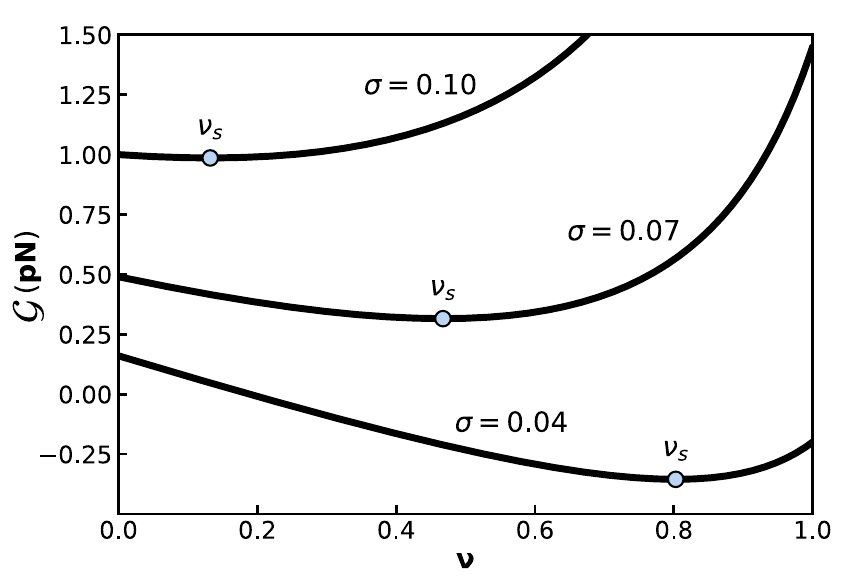}
    \caption{Plot of ${\cal G}(\nu)$ the free energy per unit length \eqref{eq:defG} 
    vs. $\nu$, the fraction of length in the stretched phase. As $\sigma$ increases the
    minimum $\nu_s=\langle \nu \rangle$ shifts towards $\nu=0$, the pure plectonemic
    phase. Fluctuations $\langle \Delta \nu^2 \rangle$ are determined by the curvature
    around the minimum ${\cal G''}(\nu_s)$, as given in Eq.~\eqref{varnu}. 
    The curvature decreases with increasing $\sigma$, implying an increase of fluctuations
    with $\sigma$. These curves correspond to $g=1$~pN, $a=500$~pN and $b=100$~pN 
    (${\cal G}$ is a free energy per unit of length, which means that it has the dimension 
    of a force and is therefore expressed in units of pN).}
    \label{fig:free_en} 
\end{figure}

While so far we have considered the asymptotic limit $L \to \infty$, one can extend 
the analysis to finite length effects. For this purpose one can start from \eqref{eq:Z} 
and perform the (Gaussian) integral on $\phi$ explicitly 
\begin{eqnarray}
    Z(\sigma,f,L) &=& 
    \int_0^1 d\nu \int_{-\infty}^{+\infty}  d\phi
    \,\,
    e^{-\beta L\nu{\cal S}(\phi,f) }
    e^{-\beta L(1-\nu) {\cal P}(\psi)}
    \nonumber\\
    &=& 
    \ldots \int_0^1 d\nu \, e^{-\beta L {\cal G}(\nu)},
    \label{eq:Zquad}
\end{eqnarray}
where the dots indicate subleading terms in $1/L$.
The free energy ${\cal G}(\nu)$ is found to be
\begin{equation}
    {\cal G}(\nu) 
    =
    -\nu g + \Xi(\nu) \, \sigma^2, 
    \label{eq:defG}
\end{equation}
which is characterized by an effective stiffness 
\begin{equation}
    \frac{1}{\Xi(\nu)} = \frac{\nu}{a} + \frac{1-\nu}{b},
    \label{eq:H_mean}
\end{equation}
given by the weighted harmonic mean of the stiffnesses of the stretched ($a$) and 
plectonemic ($b$) phases. As expected this free energy reduces to the pure state 
free energies in the two extremal cases ${\cal G}(\nu=1) = {\cal S}$ and 
${\cal G}(\nu=0) = {\cal P}$. ${\cal G}(\nu)$ is plotted in Fig.~\ref{fig:free_en} 
for three different values of $\sigma$. In the post-buckling regime 
$\sigma_s < \sigma < \sigma_p$ the fluctuations of $\nu$ are obtained from the 
second derivative of ${\cal G}(\nu)$ calculated in $\nu_s$ from which one recovers 
\eqref{varnu}. This Gaussian approximation is valid at large $L$ and breaks down 
in the vicinity of the phase boundaries $\nu_s \approx 0$ and $\nu_s \approx 1$. 
In that case the partition function \eqref{eq:Zquad} can be obtained by numerical 
integration.

\subsubsection{Rigid rod model}
\label{sec:theory_ridig_rod}

To highlight the phenomenology of the extension fluctuations we start with 
a simple model in which thermally activated bending fluctuations within 
stretched phase are neglected. We assume that this phase consists of twistable 
rigid rods aligned along the direction of the force. This corresponds to setting
$g(f)=f$ and $a(f)=a$ in Eq.~\eqref{defS}, with $a$ constant. In the pre-buckling 
regime $0 < \sigma < \sigma_s$ one obtains average and variance of $z$ from 
differentiating ${\cal S}(\sigma,f)$ with respect to $f$. The calculation gives 
$\langle z \rangle = L$ and $\langle \Delta z^2 \rangle = 0$, showing that this 
choice of stretched phase free energy indeed represents a straight rod. In the 
post-buckling regime $\sigma_s < \sigma < \sigma_p$ differentiation of 
\eqref{eq:FEquad} is simple as the only force dependence in the parameters is $g=f$
\begin{equation}
    \frac{\langle z \rangle}{L} = -\frac{d \widetilde{\cal F}}{df} 
    =
    \frac{a}{a-b} - \frac{\sigma}{\sqrt{f}}
    \sqrt\frac{ab}{a-b}.
    \label{eq:zmodA}
\end{equation}
Using \eqref{eq:sigsp_quad} one can easily verify that \eqref{eq:zmodA} 
yields $\langle z \rangle =L$ for $\sigma = \sigma_s$ and that 
$\langle z \rangle =0$ for $\sigma = \sigma_p$.
For the variance one obtains
\begin{equation}
    \frac{\langle \Delta z^2 \rangle}{L} 
    = 
    -k_B T \frac{d^2 \widetilde{\cal F}}{df^2} 
    =
    \frac{k_BT \sigma}{2f^{3/2}} \sqrt\frac{ab}{a-b} = L \langle \Delta \nu^2 \rangle,
    \label{eq:Dz2modA}
\end{equation}
where the last equality follows from \eqref{varnu}.  
As the stretched phase does not exhibit extension fluctuations, the only 
source for fluctuations in $z$ stems from the exchange of contour 
length between the two phases (fluctuations of $\nu$). We note that 
the relation
\begin{equation}
    \langle \Delta z^2 \rangle = L^2 \langle \Delta \nu^2 \rangle,
    \label{varz_rr}
\end{equation}
follows directly from the observation that the variance of $\nu$ can be
obtained from the second derivative of the free energy in $g$. This is
because $g$ and $\nu$ enter in the Boltzmann factor of \eqref{eq:Zquad} 
in the combination $g\nu$.

\subsubsection{Twistable worm-like chain (TWLC)}
\label{sec:theory_twlc}

\begin{figure*}[ht!]
    \centering\includegraphics[width=1.00\textwidth]{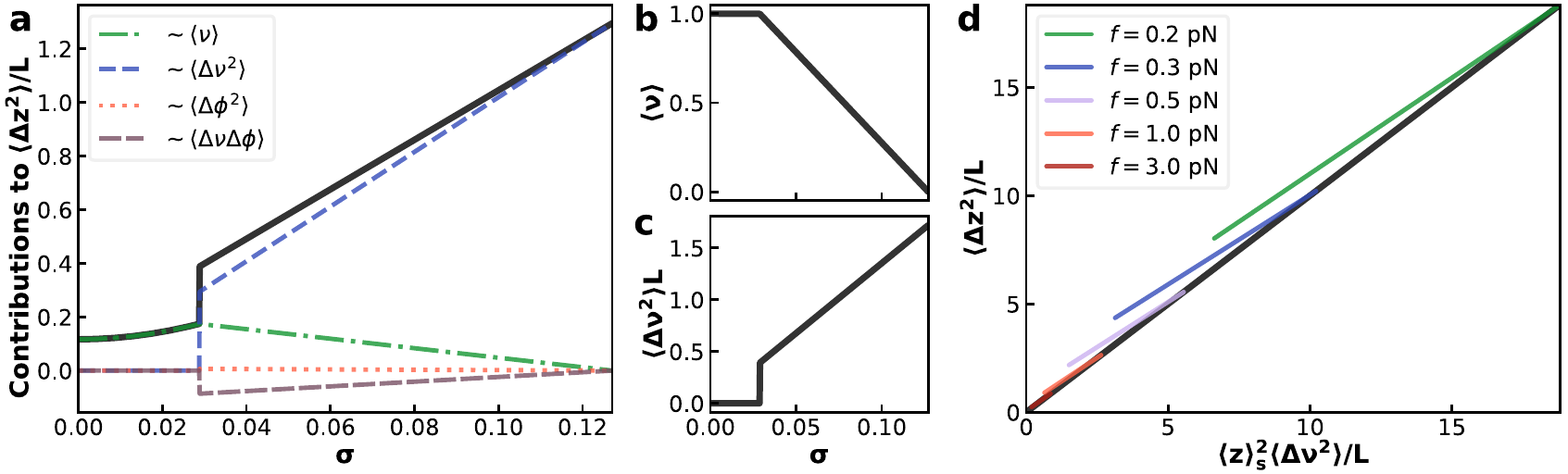}
    \caption{
    (a) Plot of the total extension variance $\vz$ vs. $\sigma$ (solid line)
    and of the four different terms (dashed, dotted and dashed-dotted colored lines) 
    on the right hand side of Eq.~\eqref{var_decomposed} with the force set to
    $f=2$~pN. The sum of these terms gives 
    $\vz$. The plot shows that the leading contribution to $\vz$ is due to the length
    exchange fluctuations, i.e. the term proportional to $\langle \Delta \nu^2 \rangle$.
    (b) Plots of $\langle \nu \rangle$ and (c) $\langle \Delta \nu^2 \rangle$ vs. $\sigma$
    for $f=2$~pN. The former decreases linearly in $\sigma$~\eqref{nu_s}, while the
    latter is discontinuous at the buckling point~\eqref{eq:gen_vnu_jump}, followed by a 
    linear increase throughout the post-buckling regime~\eqref{varnu}. (d) The colored lines 
    are plots of $\vz$ vs. $\langle z \rangle_{\cal S}^2 \, \langle \Delta \nu^2 \rangle$
    testing \eqref{final_formula} for different forces and $\sigma_s \leq \sigma \leq 
    \sigma_p$. The black line is the diagonal $x=y$ indicating perfect agreement between
    the two quantities.
    For the calculations reported in these plots we used $A=40$~nm, $C=100$~nm, $P=20$~nm 
    and $\omega_0 = 1.76\, \text{nm}^{-1}$ (corresponding to one one full turn of the 
    helix each 10.5 base pairs).}
    \label{fig:allvar-master} 
\end{figure*}

To account for the effect of bending fluctuations, we invoke
the twistable worm-like chain (TWLC) \cite{mark94,skor17,skor18,skor21}.
In this model DNA is represented as an inextensible continuous twistable rod with 
an associated bending stiffness $A$ and twist stiffness $C$.
The behavior of the TWLC within the extended phase has been studies  
for both small \cite{bouc98,bouc00} and large forces \cite{mark95,moro97,moro98}. 
Here we use the high force expansion of the stretched phase free energy, which is given 
by Eq.~\eqref{defS} with 
\begin{eqnarray}
    g(f) 
    &=& 
    f \left(1 - \sqrt{\frac{k_BT}{Af}} + \ldots \right),
    \label{gf_TWLC} 
    \\
    a(f) 
    &=& 
    \frac{C}{2} \left(1-\frac{C}{4A}\sqrt{\frac{k_BT}{Af}} + \ldots\right)
    k_BT \omega_0^2,
    \label{af_TWLC}
\end{eqnarray}
where $\omega_0 \approx 1.76$ nm$^{-1}$ is the intrinsic helical twist of DNA. 
Note, that in the limit $A \to \infty$, which amounts to suppressing bending 
fluctuations, one recovers the rigid rod model of section~\ref{sec:theory_ridig_rod}. 

Without making any particular choice for the plectoneme free energy $\pfe$, the extension 
variance $z$ can be expressed in terms of the fluctuations of $\nu$ and $\phi$. Double 
differentiation of the post-buckling free energy~\eqref{eq:FEquad} in $f$, see 
Eq.~\eqref{def:varz}, leads to the relation (details of in Appendix~\ref{app:decomposition})
\begin{eqnarray}
    \langle \Delta z^2 \rangle 
    &=& 
    k_B T L 
    \langle \nu \rangle
    ( g_{f\!f}- a_{f\!f}\sigma_s^2 ) 
    \nonumber \\
    &+& 
    L^2 \left[
    \left(g_f-a_f\sigma_s^2\right)^2\langle\dnu^2\rangle + 4a_f^2 
    \langle \nu \rangle^2 
    \sigma_s^2 \langle
    \dphi^2\rangle 
    \right.
    \nonumber \\
    &-& 
    4a_f
    \left.
    \langle \nu \rangle
    \sigma_s\left( g_f - a_f \sigma_s^2 \right) \langle \dnu\dphi \rangle
    \right],
    \label{var_decomposed}
\end{eqnarray}
where we omitted higher than second order cumulants and adopted the shorthand notation 
$g_f \equiv dg/df$ and $g_{f\!f} \equiv d^2g/df^2$. 
The correlators $\langle \dnu^2 \rangle$, $\langle \dphi^2 \rangle$ and 
$\langle \dnu\dphi \rangle$ scale as $1/L$, see Eqs.~\eqref{varnu}-\eqref{corr_dpsi}, 
while $\langle \nu \rangle$
is independent of $L$, such that all the terms on the right hand side of \eqref{var_decomposed} 
scale as $L$, i.e. they retain their relevance for large $L$.
Note that \eqref{var_decomposed} reduces to \eqref{varz_rr} in the rigid rod limit 
$g_f=1$ and $a_f=a_{f\!f}=g_{f\!f}=0$.

Eq.~\eqref{var_decomposed} decomposes $\vz$ into contributions stemming from different 
fluctuating quantities. For example, the innate stretched phase variance $\vz$ for a domain of length 
$L_s$ with supercoiling density $\sigma_s$ is
\begin{equation}
    \vz_{\cal S} = -k_B T L_s \frac{d^2 {\cal S}}{df^2} = k_BT L_s ( g_{f\!f}- a_{f\!f}\sigma^2 ).
    \label{var_stretched}
\end{equation}
At mean fractional occupancy $\langle\nu\rangle$ this domain length is $L_s=\langle\nu\rangle L$, 
which yields the first term in Eq.~\eqref{var_decomposed}. 

Evaluation of the remaining contribution requires the calculation of the correlators,
which depend on the specific functional form of the plectoneme free energy $\pfe$. 
Following prior work \cite{mark07,gao21} we invoke the harmonic form~\eqref{defP} 
with the conventional parametrization \cite{mark07}
\begin{equation}
    b=\frac 1 2 P k_B T \omega_0^2,
    \label{plec_quad_fe}
\end{equation}
where $P$ is usually referred to as the effective torsional stiffness of the plectonmic 
phase \cite{gao21}, which like $A$ and $C$ 
is expressed in units of length. A plot of $\vz$, as well as the four different contributions 
on the right hand side of Eq.~\eqref{var_decomposed}, in function of $\sigma$ for $f=2.0$~pN 
are shown in Fig.~\ref{fig:allvar-master}(a). In the pre-buckling regime, the stretched phase 
fully occupies the chain such that $\nu=1$ and $\phi=\sigma$ remain constant, leaving 
Eq.~\eqref{var_stretched} with $L_s=L$ as the only non-zero contribution to $\vz$.
This contribution decreases linearly in $\sigma$ in the post-buckling regime, reflecting the 
decrease of $\langle \nu \rangle$ with $\sigma$. 
Generally, as a consequence of the linear dependence of the free energy \eqref{eq:FEquad}
on $\sigma$ in this regime, all four contributions to $\vz$ scale linearly with $\sigma$.
We note that the term proportional to $\langle \dphi^2 \rangle$ (orange dotted line in
Fig.~\ref{fig:allvar-master}(b)) provides a very small contribution to $\vz$. 
The term proportional to the mixed correlator $\langle \dnu\dphi \rangle$ gives an overall negative
contribution to $\vz$, partially canceling the contribution of the stretched phase
fluctuations proportional to $\langle \nu \rangle$.

The analysis of Fig.~\ref{fig:allvar-master}(b) shows that the leading contribution to
$\vz$ comes from the term proportional to the phase-exchange fluctuations 
$\langle \Delta \nu^2 \rangle$.
We can rewrite this term by using the
force-extension relation for the stretched phase at supercoiled density $\sigma$ 
\begin{equation}
    \frac{\langle z \rangle_{\cal S}}{L} = -\frac{d {\cal S}}{df} = g_f - a_f \sigma^2.
\end{equation}
Using this we can then write for the leading contribution to the variance of $z$ as
\begin{equation}
    \vz \approx \langle z \rangle_{\cal S}^2 \, \langle \Delta \nu^2 \rangle,
    \label{final_formula}
\end{equation}
where $\langle z \rangle_{\cal S}$ is computed at $\sigma_s$. Figure~\ref{fig:allvar-master}(c)
shows a plot of $\vz$ vs. $\langle z \rangle_{\cal S}^2 \, \langle \Delta \nu^2 \rangle$
for different forces and $\sigma_s \leq \sigma \leq \sigma_p$ indicating good agreement
between the two terms.


\section{Twistable Worm-like chain Monte Carlo}
\label{sec:numerical}

\begin{figure}[t!]
    \centering\includegraphics[width=1\columnwidth]{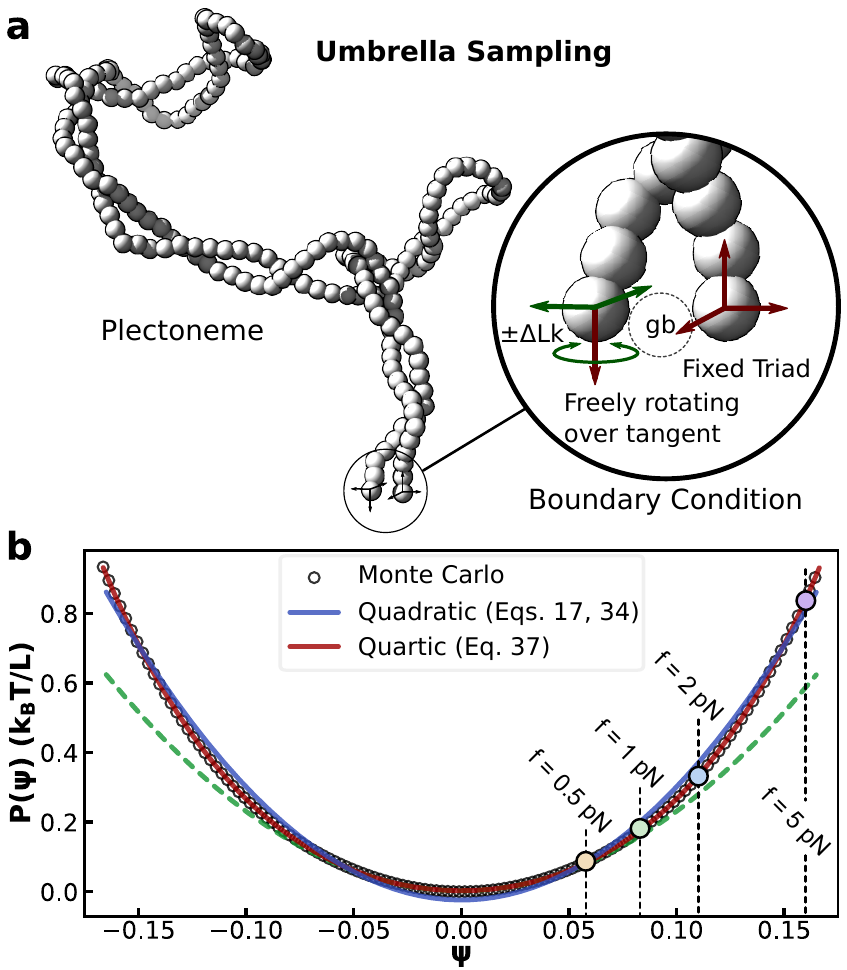}
    \caption{
    (a) Snapshot of MC simulation used to calculate the free energy of the 
    plectonemic phase $\pfe(\psi)$, via umbrella sampling. The first and last 
    beads are maintained fixed in space $8$ nm apart, with an antiparallel 
    orientation of their local tangents. The triad attached to the first 
    bead is fully prevented from rotating and that attached to
    the last bead is only permitted to rotate freely about its tangent, allowing for 
    diffusion of torsional strain into and out of the system. Between these boundary 
    beads we place demobilized ghost-bead (gb) to prevent any part of the molecule
    to cross through the open gap. By initializing each simulation with zero linking 
    number, the linking number at any snapshot is given by the accumulative rotation 
    angle of the last triad. High supercoiling densities $\sigma$
    are induced by introducing biasing torques. We perform simulations for torques 
    ranging from $1$ to $30$~pN$\cdot$nm spaced in integer steps. For each torque we sample 
    $10^7$ independent configuration for a total of $3\times10^{11}$ Monte Carlo steps.
    (b) Monte Carlo sampled free energy (circles) and approximations with quadratic and
    quartic models. Best agreement with the quadratic model 
    Eq.~\eqref{plec_quad_fe} is found for $P=20.5$~nm (solid blue line). Considering 
    the quartic model Eq.\eqref{plec_quart_fe} with $P_2=14.4$~nm and $P_4 = 524.04$~nm 
    (solid red line) improves the agreement with the MC free energy. The quadratic 
    free energy with $P=P_2$ is indicated with a green dashed line. 
    The dashed vertical lines indicate the estimated values of $\sigma_p$ (the
    average $\langle \psi \rangle$) for four different forces. Our umbrella
    sampling simulations are similar to those of Ref. \cite{klen91} performed
    on a different coarse-grained model of DNA. Deviations from harmonic behavior 
    were already noted yet not quantified in that work.}
    \label{fig:umbrella} 
\end{figure}


We tested the predictions of the two-phase model by means of Monte Carlo (MC)
simulations of the dicrete Twistable Wormlike Chain (TWLC), see e.g. \cite{vand21}. The discrete 
TWLC consists of a series of spherical beads carrying each an orthonormal 
triad of unit vectors (Fig.~\ref{fig:umbrella}a). In our simulations each coarse-grained bead 
corresponds to $10$ base pairs of DNA. From the relative orientations between two consecutive triads we compute bending 
and twist angles. The TWLC bending and twist energy is calculated from the bending and torsional 
stiffnesses given by the two parameters $A$ and $C$, respectively.
Following Rybenkov et al. \cite{rybe93} we account for both salt-concentration dependent 
electrostatic repulsion as well as steric self-avoidance by hard spheres with an effective
diameter $d_{EV}$ that exceeds the physical extension of the underlying molecule. Throughout 
this work we use $d_{EV} = 4.0$~nm, $A=40$~nm and $C=100$~nm, which were shown to produce the 
best agreement with magnetic tweezers measurements in a buffer of $150$~mM univalent salt \cite{vand21}. 
Further details on the simulation protocol can be found in supplemental material of Ref. \cite{vand21}. 


\subsection{Plectoneme Free Energy}
In a first step we utilize the MC simulations to explore the free energy  
of the plectonemic phase ${\cal P}(\psi)$ by umbrella sampling \cite{fren02}.
With appropriately imposed boundary conditions, we restrict simulations of chains of length
$L=680$~nm ($2$~kbp) to the plectonemic phase (see Fig.~\ref{fig:umbrella}b
and caption for details) and repeat the simulation for a range of biasing torques to sample 
supercoiling densities $\psi$ in the range of $-0.165 \leq \psi \leq 0.165$. 
We employ WHAM (Weighted Histogram Analysis Method) \cite{kuma92} to construct a single
unbiased histogram that combines all individual biased histograms in $\psi$.
Boltzmann-inversion then yields the sought plectoneme free energy 
${\cal P}(\psi)$, see Fig.~\ref{fig:umbrella}a. 
The best fit for the effective torsional stiffness of the plectonemic state for the 
quadratic model ${\cal P} = b \psi^2$ (with the parametrization of
Eq.~\eqref{plec_quad_fe}) over the entire sampled range is found to be $P=20.5$~nm, which 
is consistent with previous experimental measurements \cite{gao21,vand21}. However, 
closer inspection shows that MC data deviate from the quadratic model free energy.
Instead, the quartic functional form 
\begin{equation}
    \pfe(\psi) = \left(\frac{P_2}{2}\psi^2  + \frac{P_4}{4} \psi^4 \right) \omn^2 \kt,
    \label{plec_quart_fe}
\end{equation}
is found to agree significantly better with the sampled energy. 
As the supercoiling density $\psi$ is dimensionless, 
both $P_4$ and $P_2$ have the unit of a length. 
Smallest least squared differences are attained for the coefficients 
$P_2=14.4$~nm and $P_4 = 524.04$~nm. For $\psi < 0.075$ the 
relative contribution of the quartic term is less than $10\%$
(and less than $5\%$ for $\psi < 0.05$),
suggesting the quadratic model to be a reasonable approximation in this regime, albeit 
with a stiffness of $P_2$, which is about $25\%$ lower than the consensus values reported 
in the literature \cite{gao21,vand21}. 
We note that deviations from the quadratic behavior become more relevant at 
high tension as in this regime the average supercoil density of the plectonemic phase 
$\sigma_p = \langle \psi \rangle$ becomes larger ($\sigma_s$ and $\sigma_p$ grow with $f$, 
see Eqs.~\eqref{eq:sigsp_quad} and recall that $g \sim f$). 
In the high tension  post-buckling regime, it becomes important to 
include the quartic term. The vertical dashed lines in Fig.~\ref{fig:umbrella}b
show the values of $\sigma_p$ for four different forces.

It is important to note that the double tangent construction (Eq.~\ref{maxwell_FE}) 
is determined not only by the free energies $\sfe(\phi,f)$ and $\pfe(\psi)$, but 
also their local derivatives. While the global quadratic fit of the free energy may 
appear to be a reasonable approximation for ${\cal P}$ over the full range 
of $\psi$, the quartic term is necessary for a good approximation 
of the free energy  derivatives ${\cal P}_\psi$ and ${\cal P}_{\psi\psi}$
(the latter contributes to the variances \eqref{eq:gen_vnu} and 
\eqref{eq:gen_vpsi}).


\begin{figure}[t!]
    \centering\includegraphics[width=1\columnwidth]{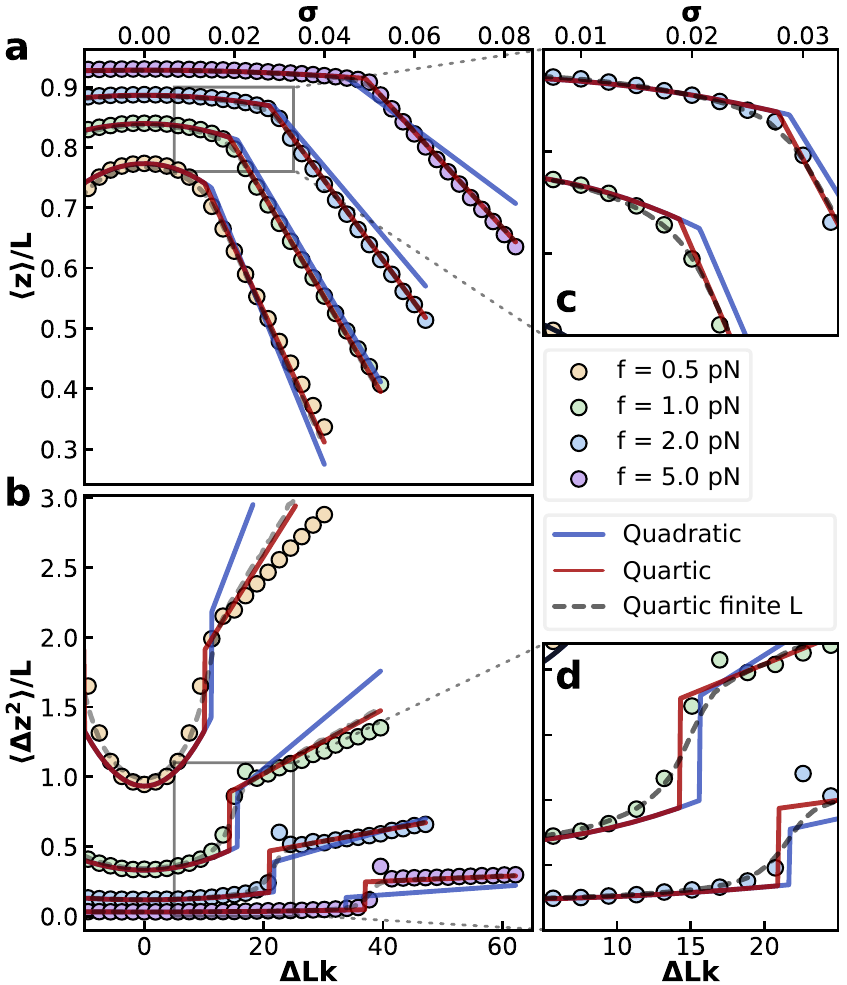}
    \caption{Rotation curves of (a) extension and (b) extension variance in function of 
    applied turns ($\dlk$ or $\sigma$) for 4 different stretching forces ($0.5$~pN, 
    $1$~pN, $2$~pN and $5$~pN).
    Colored markers show $\mz$ and $\vz$ for Monte Carlo simulations of the TWLC for a molecule
    of length $L=2692.8$~nm ($7920$~bp). For each combination of force and linking density we 
    performed $10^{11}$ Monte Carlo steps sampling the extension every $100$ steps.
    Solid blue lines give the respective values
    for the two-phase model with quadratic plectoneme free energy $\pfe$ using $P=20.5$~nm. 
    Two-phase model curves with quartic $\pfe$ using $P_2=14.4$~nm and $P_4 = 524.04$~nm are 
    shown as solid red lines. Grey solid lines show the a numerical integration of the 
    two-phase model with quartic free energy. 
    Panels (c) and (d) show a zoom-in on the buckling transition for $\mz$ and $\vz$,
    respectively. The finite length numerical calculation, with $L$ matching the 
    chain length considered in the Monte Carlo simulations, is indicated by the 
    grey dashed line.}
    \label{fig:mc_rotcurve} 
\end{figure}

\begin{figure*}[t!]
    \centering\includegraphics[width=2\columnwidth]{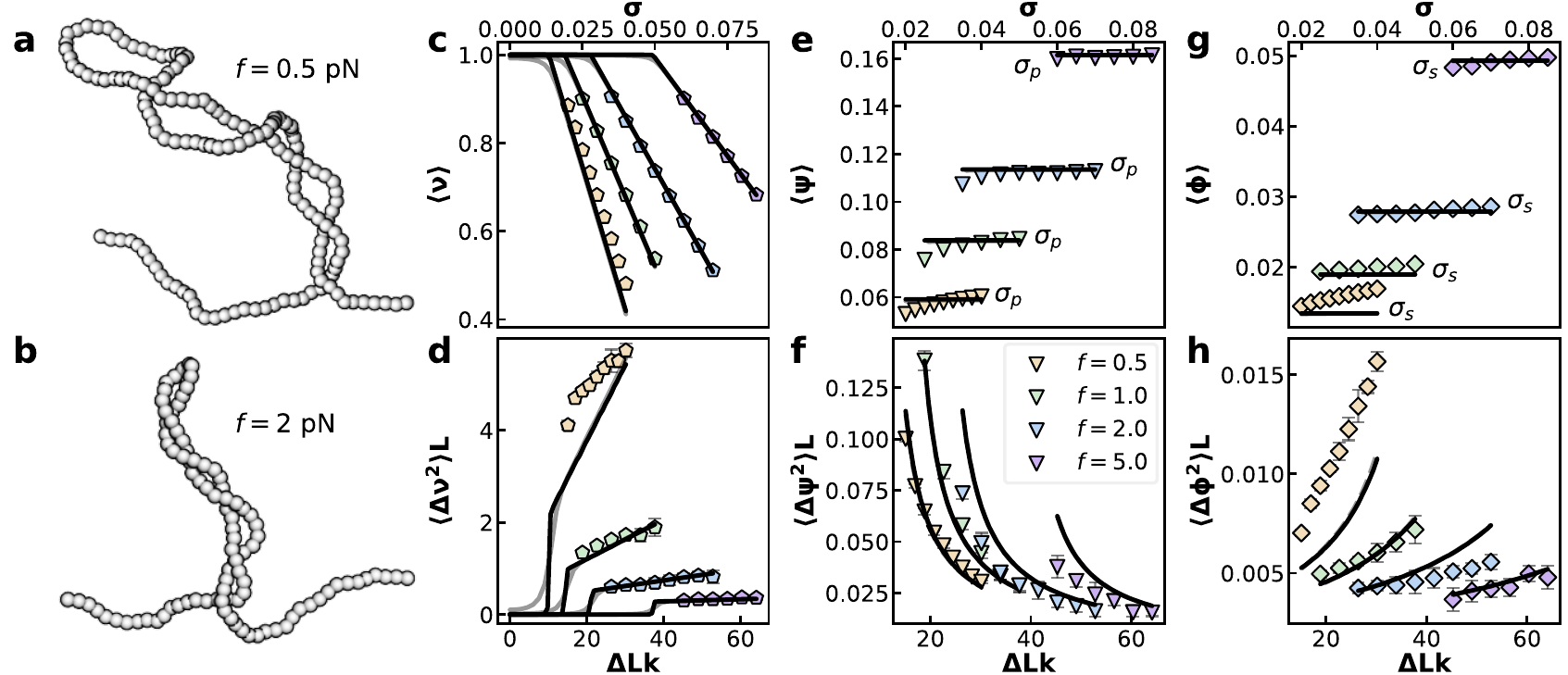}
    \caption{
    (a,b) Snapshots of MC simulations for two different forces. At higher 
    tension (b) the plectoneme is considerably tighter than at lower tension (a)
    which is reflected by a significantly larger supercoiling density $\sigma_p$.
    Using a plectoneme-detecting algorithm (Appendix~\ref{app:plecdetect}) we 
    have computed averages and variances of the fraction of length in the straight 
    phase $\nu$ (c, d), and the supercoiling density of the plectonemic phase $\psi$ 
    (e, f) and of the straight phase $\phi$ (g, h). Solid lines in c-g are the
    predictions of the two-phase model, using the quartic free energy for the
    plectonemic phase \eqref{plec_quart_fe} and the expansion \eqref{eq:g2} for
    $g(f)$. 
    }
    \label{fig:mc_tpdofs} 
\end{figure*}

\subsection{The mean extension and the extension  variance}
\label{subsec:mc-varz}
Next, we consider linear DNA molecules of length $L=2692.8$ nm ($7920$~bp), subject
to different stretching forces ($0.5$ to $5$~pN). We performed Monte 
Carlo simulations in the fixed linking number ensemble within a range of relevant 
supercoiling densities. From the sampled ensembles we obtain data of mean extension $\mz$
(Fig.~\ref{fig:mc_rotcurve}a) and extension variance $\vz$ (Fig.~\ref{fig:mc_rotcurve}b),
which we compare to the predictions of the two-phase model. For the stretched phase 
free energy we use Eq.~\eqref{defS} with $g(f)$ and $a(f)$ defined by Eqs.~\eqref{gf_TWLC} 
and \eqref{af_TWLC}. We include the next higher order term in the expansion of $g$
\begin{equation}
    g(f) = f\left( 1 - \sqrt{\frac{\kt}{Af}} + g_2 \frac{\kt}{Af} \right).
    \label{eq:g2}
\end{equation}
Using the numerically exact solution of a stretched Wormlike Chain \cite{mark95} 
we find $g_2=0.3$ for $A=40$~nm, and the room temperature value $\kt=4.1$~pN$\cdot$nm. 
The term proportional to $g_2$ contributes as an overall force-independent
constant to the stretched phase free energy, and thus does not modify the statistical
properties of this phase. However, $g_2$ is relevant in the double tangent construction
as it gives a constant relative shift to the stretched and plectoneme phase free 
energies. 

The quadratic free energy two-phase model as defined in Eq.~\eqref{plec_quad_fe} using the 
umbrella sampled full range optimization $P = 20.5$~nm yields reasonable agreement with the 
MC sampled mean extension $\mz$, albeit with visible deviations in the post-buckling slopes.
As anticipated in the discussion of the free energy, the most substantial deviations are 
observed for large forces. Large forces imply tight plectonemic coiling (large 
$\langle \psi \rangle = \sigma_p$), where the deviation of ${\cal P}$ from the quadratic 
model are most notable, see Fig.~\ref{fig:umbrella}a.
The deviations between theory and simulations are almost entirely resolved by invoking the 
quartic model \eqref{plec_quart_fe}, which yields excellent agreement with the mean extension 
of the MC data across the full range of forces. 

In spite of the success of the quartic model for the mean extension, the variance 
reveals the existence of remnant shortcomings. While the general
features are quite accurately reproduced and the high force agreement is rather  
convincing, the slope of the low force variance significantly overestimates the 
MC data. Nonetheless, the quartic free energy constitutes a compelling improvement
to the quadratic model.

We also performed a numerical integration of the full partition function. 
This reveals finite $L$ effects that are not captured by the free energy minimization 
approach \eqref{eq:Z}, which is valid in the thermodynamic limit. 
The difference between the
two calculations is mostly visible in the vicinity of the phase boundary $\sigma=\sigma_s$
(see Fig.~\ref{fig:mc_rotcurve}c,d).
Here the Gaussian approximation predicts a jump of $\vz$, while the numerical 
integration smoothly connects the pre- and post-buckling regimes, and matches
the simulation data in the pre-buckling regime.


\subsection{Two-phase model degrees of freedom}
\label{subsec:mc_2phase}
Inspired by previous work \cite{volo92,liu08a,coro18}, we developed an algorithm to detect
the regions occupied by the plectonemic phase in simulation generated snapshots 
(see Appendix~\ref{app:plecdetect} for details). With this algorithm we calculated
the total fraction of stretched phase $\nu$, and the supercoiling densities
of the stretched and plectonemic phases, $\phi$ and $\psi$. 
We compared these results with the prediction of the two-phase model, employing 
the quartic model (Eq.~\eqref{plec_quart_fe}) for the plectoneme free energy with the 
parameters obtained from umbrella sampling.
The averages $\langle \nu \rangle$, $\langle \psi \rangle$ and $\langle \phi \rangle$
convincingly follow the prediction of the two-phase model, at least for large forces 
(Figs.~\ref{fig:mc_tpdofs}a, c and e).
We note that $\langle \nu \rangle$ is linear in $\sigma$, as expected from Eq.~\eqref{nu_s}.
Moreover, $\langle \phi \rangle = \sigma_s$ and $\langle \psi \rangle = \sigma_p$ are 
approximately independent of $\sigma$. Deviations are apparent for the smallest force ($0.5$~pN) 
and close to the buckling point $\sigma \approx \sigma_p$, where finite length effects
are of relevance. 
The variances $\langle \dnu^2 \rangle$, $\langle \dpsi^2 \rangle$ and $\langle \dphi^2 \rangle$
show stronger deviations from the theory as compared to the averages, especially at the smallest
force analyzed $f=0.5$~pN.
Considering the inherent difficulty to determine phase-boundaries, and the sensitivity of higher 
moments to algorithmic noise, we observe satisfactory agreement between theory and simulations
for the variances of $\nu$, $\psi$ and $\phi$  (Figs.~\ref{fig:mc_tpdofs}(b),(d) and (f)). 
For all forces the two-phase model theory qualitatively captures the general 
trends of the fluctuations. Quantitative agreement is once again limited to the higher 
force regime $f \gtrsim 1$~pN.


\begin{figure}[t]
    \centering\includegraphics[width=1\columnwidth]{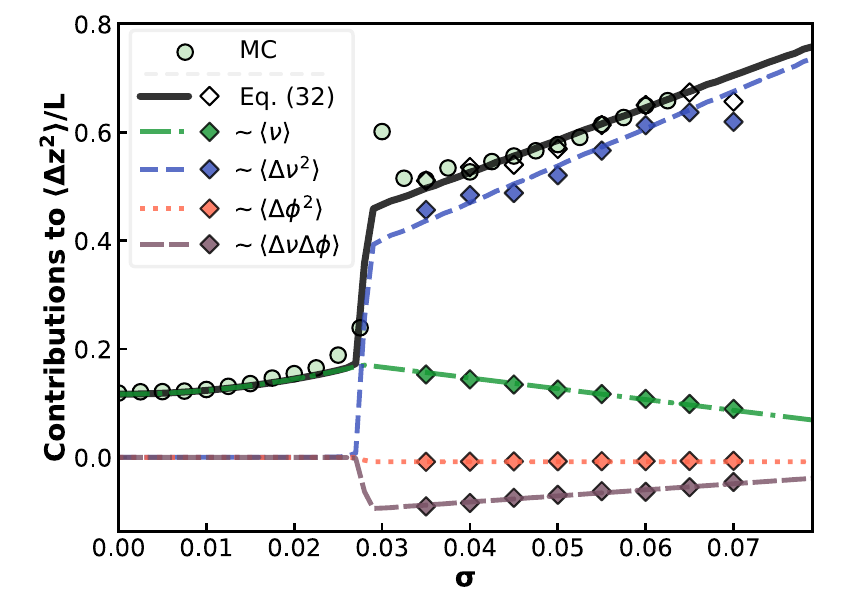}
    \caption{
    Comparison of the extension variance $\vz$ for $f=2$~pN (black solid 
    line: theory with quartic $\pfe$; round symbols: TWLC MC) to the four
    different contributions according to Eq.~\eqref{var_decomposed} (blue: term 
    $\sim\vnu$; green: term $\sim\mnu$; orange: term $\sim\vphi$; brown: term 
    $\sim\ccnuphi$). The numerical calculations with quartic $\pfe$ are shown as 
    lines and MC data as diamonds. White diamonds represent the sum of the MC 
    extracted contributions.}
    \label{fig:mc_vzcontr} 
\end{figure}

\subsection{Contributions to extension variance}
\label{subsec:mc_vzcontr}
Finally, we verify if the conclusions of Section~\ref{sec:theory_twlc} that were obtained
for the two-phase model with quadratic free energies still hold for a quartic plectoneme 
free energy and for the TWLC Monte Carlo simulations. The expectation values for $\mnu$, 
$\vnu$, $\vphi$ and $\ccnuphi$, calculated numerically for the for quartic $\pfe$
and deduced from TWLC Monte Carlo simulations as shown in Fig.~\ref{fig:mc_tpdofs}, 
allow us to once again decompose the extension variance into various contributions 
according to Eq.~\eqref{var_decomposed}. Note, that this equation remains valid as long 
as the stretched phase free energy is reasonably well approximated by the stretching 
free energy~\eqref{gf_TWLC} and twist free energy~\eqref{af_TWLC}. Given the excellent 
agreement between the theoretical and simulated $\mz$ and $\vz$ in the pre-buckling 
regime (see Fig.~\ref{fig:mc_rotcurve}) we consider this requirement satisfied. 
The differences between the quadratic and quartic $\pfe$ are fully contained in the 
the correlators (see Eq.~\eqref{var_decomposed}).

In Fig.~\ref{fig:mc_vzcontr} the extension variance of the TWLC Monte Carlo 
simulations for $f=2$~pN is compared to the contributions stemming from $\mnu$, $\vnu$, $\vphi$ 
and $\ccnuphi$ according to Eq.~\eqref{var_decomposed}. 
Just as in the case of the quadratic free energy, the extension variance $\vz$ is 
dominated by phase exchange fluctuations $\vnu$, especially for large supercoiling 
densities, where the system shows a significant occupancy of the plectonemic phase.


\section{Conclusion}
\label{sec:conclusion}
In this paper we discussed the properties of end-point fluctuations of linear 
stretched and overtwisted DNA. This is the typical setup of single molecule MT 
experiments in which a magnetic field is used to apply a stretching force $f$ and 
to maintain kbp-long DNA molecule at fixed linking number. We based our analysis 
on a two-phase model \cite{mark07}, which describes DNA buckling as a thermodynamic 
first order transition. 

In the pre-buckling regime, both mean extension $\mz$ as well
as extension fluctuations, characterized by the variance $\vz$, are 
controlled by the behavior of the stretched phase. Past the buckling point, 
the DNA consists of coexisting stretched and plectonemic domains.
The average length of the plectonemic phase increases with increasing 
supercoiling density $\sigma$, while the stretched phase decreases, leading to
a global decrease of $\langle z \rangle$. 
The two-phase model indicates that several terms contribute to $\vz$, representing 
distinct fluctuation mechanisms, see Eq.~\eqref{var_decomposed}. While $\vz$ linearly 
increases with $\sigma$ at post-buckling, the contribution of extension fluctuations 
of the stretched phase (term proportional to $\nu$ in \eqref{var_decomposed}) 
decreases, as this phase decreases in length. We show that of all the contributions 
to $\vz$, phase-exchange fluctuations, e.g. the transient exchange 
of lengths between the stretched and the plectonemic phases through which the 
plectonemes shrink and grow, is the dominant one.
Equation~\eqref{final_formula}, which approximates \eqref{var_decomposed}, 
summarizes the dominance of the length exchange mechanism.
The remaining contributions to the extension variance $\vz$ are due to fluctuations
of the supercoiling densities of the stretched and plectonemic phases and of mixed 
terms coupling stretched phase length and supercoiling density. Although the overall 
contribution of these terms to $\vz$ is minor, we note a perhaps curious cancellation 
between a (positive) stretched phase fluctuation term and a (negative) mixed 
correlator term. These two contributions are 
indicated as $\sim \langle \nu \rangle$ and $\sim \langle \dphi \dnu \rangle$ in 
Figs.~\ref{fig:allvar-master}a and \ref{fig:mc_vzcontr}. The contribution to $\vz$ 
due to the fluctuations of stretched phase supercoil density (term indicated with 
$\sim \langle \dphi^2 \rangle$ in Figs.~\ref{fig:allvar-master}a and 
\ref{fig:mc_vzcontr}) is generally small.

We corroborate these theoretical predictions with Monte Carlo simulations of the discrete TWLC.
In a first step we ascertain the most appropriate form for the plectonemic free energy
$\pfe$ by umbrella sampling. We find that the quadratic 
free energy~\eqref{defP} is a good approximation only for plectoneme supercoiling densities 
$\sigma_p$ smaller than roughly $0.05$. Beyond that, the quartic~\eqref{plec_quart_fe} is 
found to yield satisfactory agreement. For large forces, the use of the quartic $\pfe$
is found to significantly increase the agreement between theoretical extension and extension 
variance curves and Monte Carlo sampled data as compared to the previously used
quadratic \cite{mark07,vand21}. 
Our analysis suggests that the two-phase model captures the phenomenology of the 
fluctuations in the post-buckling regime. Discrepancies observed at small forces 
($f\leq1$~pN) are likely due to a breakdown of a two-phase model description.
At small forces, both plectonemes and stretched parts are loose and
are likely to be separated by a smooth and extended interfacial region.
This leads to a difficulty in sharply separating plectonemes from stretched domains. 
In spite of these deviations, the overall qualitative picture holds also at small forces.

Summarizing, we believe that an in-depth understanding of the nature of equilibrium 
fluctuations of stretched supercoiled DNA is important, since in the cell, as well 
as in {\sl in-vitro} experiments, DNA subject to strong thermal fluctuations. 
Moreover, fluctuations can carry information about processes that do not affect 
average quantities, as was illustrated in a recent study of protein binding on
supercoiling DNA \cite{vand21}.


\begin{acknowledgments}
Discussions with Pauline Kolbeck, Jan Lipfert, Midas Segers and 
Willem Vanderlinden are gratefully acknowledged. 
ES acknowledges financial support from Fonds Wetenschappelijk 
Onderzoek (FWO) Grant 1SB4219N.
\end{acknowledgments}


\appendix


\section{Gaussian fluctuations of $\nu$ and $\phi$}
\label{app:fluctuations}
Here, we provide additional details on the calculations of the fluctuations 
in $\nu$, $\phi$ and $\psi$ given in Eqs.~\eqref{eq:gen_vnu}, \eqref{eq:gen_vphi}, 
\eqref{eq:gen_dnudphi} and \eqref{eq:gen_vpsi}. We use the same notation as in 
the main text: subscripts denote differentiation with respect to the indicated 
variable followed by evaluation at the minimum. For example
\begin{eqnarray}
    \pfe_\psi = \left.\frac{d \pfe}{d\psi}\right|_{\psi=\sigma_p}, 
    &\quad&
    \psi_\nu = \left.\frac{d \psi}{d\nu}\right|_{\phi=\sigma_s,\nu=\nu_s}.
\end{eqnarray}
Using the expression for the free energy \eqref{eq:twophase_fe} we find
for the second derivative in $\nu$
\begin{eqnarray}
    \ffe_{\nu\nu} &=& -2 \pfe_\psi \psi_\nu + (1-\nu_s) 
    \left( \pfe_\psi \psi_{\nu\nu} +  \pfe_{\psi\psi} \psi_\nu^2 \right)
    \nonumber \\
    &=& 
    (1-\nu_s) \pfe_{\psi\psi} \psi_\nu^2 = 
    \frac{(\sigma_p-\sigma_s)^2}{1-\nu_s} 
    \pfe_{\psi\psi},
    \label{Fnunu}
\end{eqnarray}
where we have used the following relations
\begin{eqnarray}
    \psi &=& \frac{\sigma - \phi}{1-\nu} - \phi, \\
    \psi_\nu &=& \frac{\sigma-\sigma_s}{(1-\nu_s)^2} = \frac{\sigma_p-\sigma_s}{1-\nu_s}, \\
    \psi_{\nu\nu} &=& 2 \frac{\sigma-\sigma_s}{(1-\nu_s)^3} = \frac{2\psi_\nu}{1-\nu_s}.
\end{eqnarray}
Likewise, we find for the other derivatives
\begin{align}
    \ffe_{\phi\phi}
    &=
    \frac{\nu_s}{1-\nu_s}\left[(1-\nu_s)\sfe_{\phi\phi} + \nu_s \pfe_{\psi\psi} \right], 
    \label{Fphiphi}
    \\
    {\cal F}_{\phi\nu}
    &=
    \frac{\nu_s}{1-\nu_s} (\sigma_p-\sigma_s) \pfe_{\psi\psi}.
    \label{Fpsipsi}
\end{align}
Equipartition then stipulates that fluctuations in $\phi$ and $\nu$ are obtained from 
the inversion of the Hessian matrix. This gives
\begin{eqnarray}
    \beta L \langle \Delta \nu^2 \rangle = 
    \frac{{\cal F}_{\phi\phi}}{{\cal F}_{\nu\nu}{\cal F}_{\phi\phi}-{\cal F}_{\nu\phi}^2},
    \\
    \beta L \langle \Delta \phi^2 \rangle = 
    \frac{{\cal F}_{\nu\nu}}{{\cal F}_{\nu\nu}{\cal F}_{\phi\phi}-{\cal F}_{\nu\phi}^2},
    \\
    \beta L \langle \Delta \nu \Delta \phi\rangle = -
    \frac{{\cal F}_{\nu\phi}}{{\cal F}_{\nu\nu}{\cal F}_{\phi\phi}-{\cal F}_{\nu\phi}^2}.
\end{eqnarray}

Replacing \eqref{Fnunu}, \eqref{Fphiphi} and \eqref{Fpsipsi} in the previous relations, 
we get Eqs.~\eqref{eq:gen_vnu}, \eqref{eq:gen_vphi} and \eqref{eq:gen_dnudphi}.


\section{Derivation of $\vz$ for the TWLC}
\label{app:decomposition}
The free energy per unit length of the quadratic model in the post-buckling regime 
\eqref{eq:FEquad} depends on the force $f$ via the variables $a(f)$ and $g(f)$. 
Differentiation with respect to $f$ can be expressed via the partial derivatives 
with respect to $a$ and $g$ as
\begin{equation}
    {\cal F}_f =  {\cal F}_g \, g_f + {\cal F}_a \, a_f,
\end{equation}
where we use the notation $\Psi_f \equiv d\Psi/df$ for the total derivative in $f$ and 
$\Psi_a \equiv \partial \Psi/\partial a$ and $\Psi_g \equiv \partial \Psi/\partial g$
for partial derivatives in $a$ and $g$. The second derivative in $f$ gives the variance
in $z$ which can then be written as
\begin{eqnarray}
    &\beta \frac{\vz}{L} = -{\cal F}_{f\!f} = - \left({\cal F}_g \, g_{f\!f} + {\cal F}_a \, a_{f\!f} +  
    {\cal F}_{gg} \, g_f^2 +  {\cal F}_{aa} \, a_f^2
    \right. \nonumber \\
    &\left.
     + 2 {\cal F}_{ag} \, a_f g_f \right)
    =
    \langle \nu \rangle g_{f\!f} - \langle \nu \phi^2 \rangle a_{f\!f} 
            + \beta L \left[
    \langle \Delta \nu^2 \rangle g_f^2 + \right.
    \nonumber\\        
    &
    \left. \langle \Delta (\nu \phi^2)^2 \rangle a_f^2
    - 2 \left( \langle \nu^2 \phi^2 \rangle - \langle \nu \rangle 
     \langle \nu \phi^2\rangle \right) a_f g_f
    \right]
    \label{Fff},
\end{eqnarray}
where we have expressed derivatives with respect to $a$ and $g$ as correlators of combinations 
of variables $\nu$ and $\phi$ using
\begin{eqnarray}
     {\cal F}_{g} 
     &=& 
     - \langle \nu \rangle, \quad
     {\cal F}_{gg} = - \beta  L \langle \Delta \nu ^2 \rangle, \\
     {\cal F}_{a} 
     &=&  
     \langle \nu \phi^2 \rangle, \quad
     {\cal F}_{aa} = - \beta L \langle \Delta (\nu \phi^2)^2 \rangle, \\
     {\cal F}_{ag} 
     &=&  
     \beta L \left( \langle \nu^2 \phi^2 \rangle - \langle \nu \rangle 
     \langle \nu \phi^2\rangle \right),
     \label{last}
\end{eqnarray}
which can be obtained from the form of the free energy. These correlators can be further
expanded to lowest order in fluctuations around the averages $\Delta \phi = \phi -\sigma_s$ 
and $\Delta \nu = \nu - \nu_s$ (recall that $\nu_s = \langle \nu \rangle$ and $\sigma_s = 
\langle \phi \rangle$). 

For example, for the term $\langle \nu \phi^2 \rangle$ this expansion gives
\begin{eqnarray}
    \langle \nu \phi^2 \rangle  
    &\approx& 
    \nu_s\sigma_s^2 + \nu_s \langle \dphi^2 \rangle + 2 \sigma_s \langle \dnu\dphi\rangle. 
\end{eqnarray}
Similar relations can be obtained for $\langle \nu^2 \phi^2 \rangle$ and $\langle\Delta(\nu\phi^2)^2\rangle$. 
Substituting these in \eqref{Fff}, we can express the variance in $z$ as a function of the correlators 
$\langle\dnu^2\rangle$, $\langle  \dphi^2 \rangle$ and $\langle \dnu\dphi \rangle$ which gives
Eq.~\eqref{var_decomposed}.


\section{Detecting Plectonemes}
\label{app:plecdetect}
\begin{figure}[t!]
    \centering\includegraphics[width=1\columnwidth]{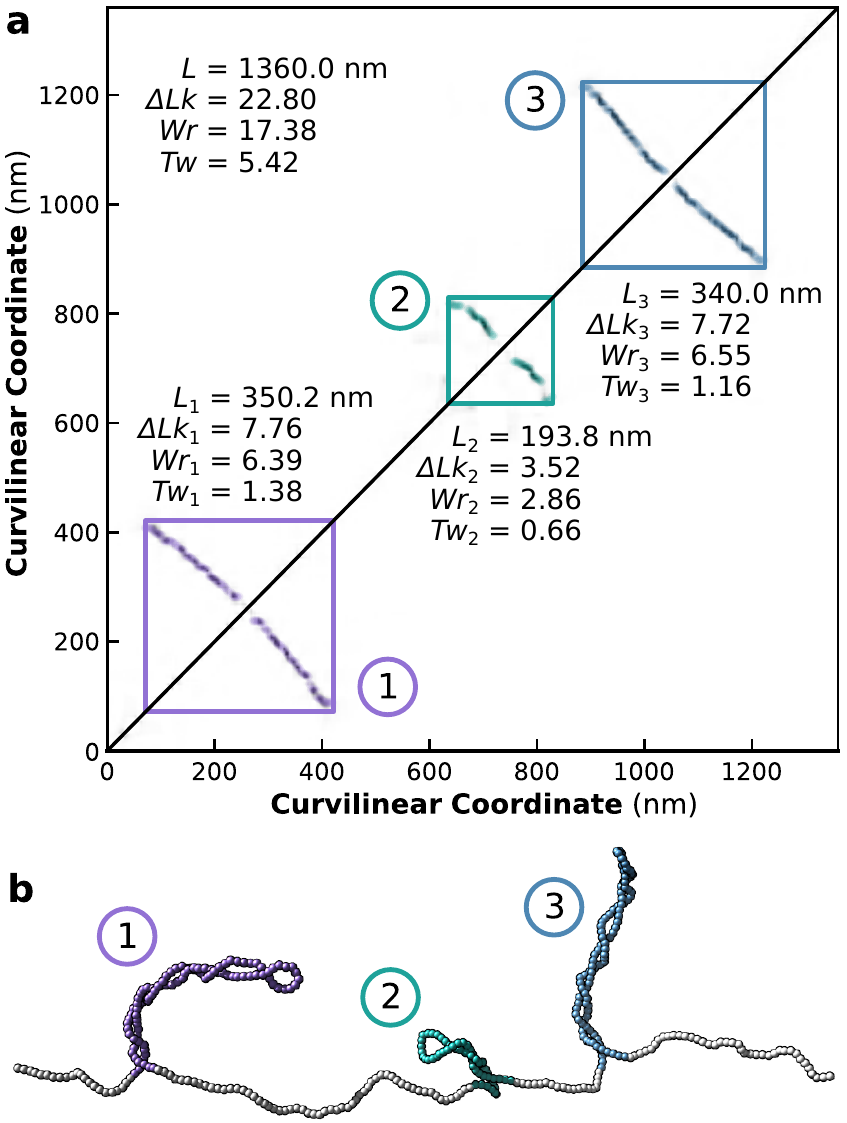}
    \caption{
    Illustration of the algorithm used to identify plectonemes in Monte Carlo
    generated snapshots. (a) Writhe map corresponding to the configuration shown
    in (b). The three plectonemes of panel (b) are highlighted by rectangles 
    with corners $(i,i)$, $(i,j)$, $(j,i)$ and $(j,j)$ for a plectoneme with 
    entry and exit indices $i$ and $j$, respectively. In the figure these 
    indices have been converted into curvilinear coordinates. Writhe and twist
    can be calculated separately for each plectoneme, which allows for the 
    calculation of the plectoneme linking number. 
    }
    \label{fig:plecdetect} 
\end{figure}

Here we outline the algorithm used to detect plectonemic regions in
simulation generated configurations. Two different characteristics have been used
in the past to identify plectonemes. For one, plectonemic coiling induces juxtaposition 
between sites that are far away along the DNA contour. Therefore, various studies,
utilized a distance map (or contact map) to identify regions of high segment 
proximity \cite{mate15,coro18,desa20}. 
Alternatively, one can identify plectonemes as regions of high writhe density
\cite{volo92,liu08a}. Writhe (defined below) is a measure of the amount of coiling
of a closed curve \cite{full71}, and plectonemes are characterized by a considerably 
larger writhe density as compared to the stretched phase (see Fig.~\ref{fig:mc_tpdofs} 
and \cite{liu08a,mark15}).

In this work, we used the latter property to identify plectonemes. Since our simulations are composed of a series 
of $N-1$ straight segments $\mathbf{s}_i$ connecting consecutive beads, the double integral definition of writhe
\cite{full71} may be decomposed into $(N-1)^2$ pairwise contributions~\footnote{We remark that writhe is technically 
only defined for a closed space curve, which in the context of 
stretched linear DNA is usually resolved by introducing a virtual closure at either infinity or through a large 
arc \cite{volo97,chou14}. We ignore these closure contributions, since they are irrelevant for the detection of plectonemes.}
\begin{equation}
    \textrm{Wr}  = 
    \frac{1}{4\pi} \sum_{i=1}^N\sum_{j=1}^N \int_{\mathbf{s}_i}\int_{\mathbf{s}_j} \frac{\left(d\mathbf{r}_j 
    \times d\mathbf{r}_i\right)\cdot\mathbf{r}_{ij}}{r_{ij}^3} = 
    \sum_{i=1}^N\sum_{j=1}^N w_{ij}.
\end{equation}
The elements $w_{ij}$ may then be viewed as analogous to the entries of a contact map. In practice, 
we calculate those elements $w_{ij}$ by analytically carrying out the double integral over the two 
straight segments $\mathbf{s}_i$ and $\mathbf{s}_j$ of length $a$ ($3.4$~nm in our case). For details 
on this calculation see method $1b$ from Ref.~\cite{klen00}. Note, that in the writhe map, plectonemes 
have much more contrast as compared to the proximity map ~\cite{liu08a}. 
An example of a writhe map for a Monte Carlo generated configuration of length $L=1360$~nm ($4000$~bp 
or $400$ segments) containing 3 plectonemes is shown in Fig.~\ref{fig:plecdetect}. 
The high contrast black bands trace segment pairs $(s_i,s_j)$, where the $i$-th and $j$-the segment 
are on opposing strands along the superhelix. Towards the diagonal, as $i$ comes close to $j$, the 
trace approaches the plectoneme end-loop. Conversely, the outermost segments, indicated by the 
upper-left and lower-right corner of the rectangles drawn around the plectonemic region, mark the 
entry points of the plectonemes. 

We trace plectonemes by setting a force dependent cutoff writhe-density $\chi_{\mathrm{min}}$, 
which is typically chosen, somewhere in-between the expected stretched and plectonemic phase 
supercoiling densities $\sigma_s$ and $\sigma_p$. Such choice can be made more quantitative by 
also considering the twist-density in the chain, which can either be directly calculated from 
the given snapshots or from the stretched phase theory \cite{moro98,nomi20_thesis,eman13} 
\begin{equation}
    \sigma_{Tw} = \left(1-\frac{C}{4A}\sqrt{\frac{\kt}{Af}}\right)\sigma,
\end{equation}
as twist is equilibrated over both phases. We then identify all indices $i$ for which the 
total involved writhe density is at least $\chi_{\mathrm{min}}$, i.e.
\begin{equation}
    W_i \equiv \sum_{j=1}^{N-1} w_{i,j} >= \frac{a \omega_0}{2\pi}\chi_{\mathrm{min}}.
\end{equation}
Neighbors among these remaining segments, say at indices $k$ and $l$ are then connected, if 
the sum of the intermediate contributions of $W_i$, including those at $k$ and $l$, still 
exceeds the minimum writhe density, i.e. if
\begin{equation}
    \frac{1}{l-k+1}\sum_{j=k}^l W_j >= \frac{a \omega_0}{2\pi}\chi_{\mathrm{min}}.
\end{equation}
This generates connected regions tracing one strand along a superhelical branch. The branch itself, 
can be traced by identifying for every $i$ the index $j$ for which the writhe contribution $|w_{ij}|$ is largest.
Full plectoneme branches are finally identified by finding the pairs of strands which map into each other by inverting 
the respective indices, which is equivalent to transposing the points relative to the writhe map. Certain care has to 
be taken when tracing the branches, not to pick up contributions stemming from outside the current branch, which could 
in some cases lead to an erroneous assignment of a part of a stretched domain to a plectoneme. An exhaustive 
discussion of the algorithm is beyond the scope of this work. The final traces found for each plectoneme
are indicated as colored lines in Fig.~\ref{fig:plecdetect}.

Once the plectonemic domains are identified the contained writhe can be extracted from the writhe map. The twist 
can either be taken directly from the simulation, if local twist strains are consisted, or in the case of 
equilibrated twist simulations~\cite{klen91,yang00,lepa15} from the total twist $Tw = \Delta Lk - Wr$ of the 
entire snapshot. Summation of writhe and twist contained in the plectonemic domains yields the total linking number
in the plectonemic phase, which together with the number of segments in in these domains yields the supercoiling 
density $\psi$. Completely analogously is the calculation of the the stretched phase supercoiling density $\phi$. 
Finally, the fraction of segments not contained in plectonemic domains gives the fractional occupancy of the 
stretched phase $\nu$.


\bibliography{main}

\begin{thebibliography}{52}%
\makeatletter
\providecommand \@ifxundefined [1]{%
 \@ifx{#1\undefined}
}%
\providecommand \@ifnum [1]{%
 \ifnum #1\expandafter \@firstoftwo
 \else \expandafter \@secondoftwo
 \fi
}%
\providecommand \@ifx [1]{%
 \ifx #1\expandafter \@firstoftwo
 \else \expandafter \@secondoftwo
 \fi
}%
\providecommand \natexlab [1]{#1}%
\providecommand \enquote  [1]{``#1''}%
\providecommand \bibnamefont  [1]{#1}%
\providecommand \bibfnamefont [1]{#1}%
\providecommand \citenamefont [1]{#1}%
\providecommand \href@noop [0]{\@secondoftwo}%
\providecommand \href [0]{\begingroup \@sanitize@url \@href}%
\providecommand \@href[1]{\@@startlink{#1}\@@href}%
\providecommand \@@href[1]{\endgroup#1\@@endlink}%
\providecommand \@sanitize@url [0]{\catcode `\\12\catcode `\$12\catcode
  `\&12\catcode `\#12\catcode `\^12\catcode `\_12\catcode `\%12\relax}%
\providecommand \@@startlink[1]{}%
\providecommand \@@endlink[0]{}%
\providecommand \url  [0]{\begingroup\@sanitize@url \@url }%
\providecommand \@url [1]{\endgroup\@href {#1}{\urlprefix }}%
\providecommand \urlprefix  [0]{URL }%
\providecommand \Eprint [0]{\href }%
\providecommand \doibase [0]{https://doi.org/}%
\providecommand \selectlanguage [0]{\@gobble}%
\providecommand \bibinfo  [0]{\@secondoftwo}%
\providecommand \bibfield  [0]{\@secondoftwo}%
\providecommand \translation [1]{[#1]}%
\providecommand \BibitemOpen [0]{}%
\providecommand \bibitemStop [0]{}%
\providecommand \bibitemNoStop [0]{.\EOS\space}%
\providecommand \EOS [0]{\spacefactor3000\relax}%
\providecommand \BibitemShut  [1]{\csname bibitem#1\endcsname}%
\let\auto@bib@innerbib\@empty
\bibitem [{\citenamefont {Vanderlinden}\ \emph {et~al.}(2019)\citenamefont
  {Vanderlinden}, \citenamefont {Brouns}, \citenamefont {Walker}, \citenamefont
  {Kolbeck}, \citenamefont {Milles}, \citenamefont {Ott}, \citenamefont
  {Nickels}, \citenamefont {Debyser},\ and\ \citenamefont {Lipfert}}]{vand19}%
  \BibitemOpen
  \bibfield  {author} {\bibinfo {author} {\bibfnamefont {W.}~\bibnamefont
  {Vanderlinden}}, \bibinfo {author} {\bibfnamefont {T.}~\bibnamefont
  {Brouns}}, \bibinfo {author} {\bibfnamefont {P.~U.}\ \bibnamefont {Walker}},
  \bibinfo {author} {\bibfnamefont {P.~J.}\ \bibnamefont {Kolbeck}}, \bibinfo
  {author} {\bibfnamefont {L.~F.}\ \bibnamefont {Milles}}, \bibinfo {author}
  {\bibfnamefont {W.}~\bibnamefont {Ott}}, \bibinfo {author} {\bibfnamefont
  {P.~C.}\ \bibnamefont {Nickels}}, \bibinfo {author} {\bibfnamefont
  {Z.}~\bibnamefont {Debyser}},\ and\ \bibinfo {author} {\bibfnamefont
  {J.}~\bibnamefont {Lipfert}},\ }\bibfield  {title} {\bibinfo {title} {The
  free energy landscape of retroviral integration},\ }\href
  {https://doi.org/https://doi.org/10.1038/s41467-019-12649-w} {\bibfield
  {journal} {\bibinfo  {journal} {Nature Comm.}\ }\textbf {\bibinfo {volume}
  {10}},\ \bibinfo {pages} {1} (\bibinfo {year} {2019})}\BibitemShut {NoStop}%
\bibitem [{\citenamefont {Yan}\ \emph {et~al.}(2018{\natexlab{a}})\citenamefont
  {Yan}, \citenamefont {Ding}, \citenamefont {Leng}, \citenamefont {Dunlap},\
  and\ \citenamefont {Finzi}}]{yan18a}%
  \BibitemOpen
  \bibfield  {author} {\bibinfo {author} {\bibfnamefont {Y.}~\bibnamefont
  {Yan}}, \bibinfo {author} {\bibfnamefont {Y.}~\bibnamefont {Ding}}, \bibinfo
  {author} {\bibfnamefont {F.}~\bibnamefont {Leng}}, \bibinfo {author}
  {\bibfnamefont {D.}~\bibnamefont {Dunlap}},\ and\ \bibinfo {author}
  {\bibfnamefont {L.}~\bibnamefont {Finzi}},\ }\bibfield  {title} {\bibinfo
  {title} {Protein-mediated loops in supercoiled {DNA} create large topological
  domains},\ }\href {https://doi.org/10.1093/nar/gky153} {\bibfield  {journal}
  {\bibinfo  {journal} {Nucl. Acids Res.}\ }\textbf {\bibinfo {volume} {46}},\
  \bibinfo {pages} {4417} (\bibinfo {year} {2018}{\natexlab{a}})}\BibitemShut
  {NoStop}%
\bibitem [{\citenamefont {Yan}\ \emph {et~al.}(2018{\natexlab{b}})\citenamefont
  {Yan}, \citenamefont {Leng}, \citenamefont {Finzi},\ and\ \citenamefont
  {Dunlap}}]{yan18b}%
  \BibitemOpen
  \bibfield  {author} {\bibinfo {author} {\bibfnamefont {Y.}~\bibnamefont
  {Yan}}, \bibinfo {author} {\bibfnamefont {F.}~\bibnamefont {Leng}}, \bibinfo
  {author} {\bibfnamefont {L.}~\bibnamefont {Finzi}},\ and\ \bibinfo {author}
  {\bibfnamefont {D.}~\bibnamefont {Dunlap}},\ }\bibfield  {title} {\bibinfo
  {title} {Protein-mediated looping of {DNA} under tension requires
  supercoiling},\ }\href {https://doi.org/10.1093/nar/gky021} {\bibfield
  {journal} {\bibinfo  {journal} {Nucl. Acids Res.}\ }\textbf {\bibinfo
  {volume} {46}},\ \bibinfo {pages} {2370} (\bibinfo {year}
  {2018}{\natexlab{b}})}\BibitemShut {NoStop}%
\bibitem [{\citenamefont {Yan}\ \emph {et~al.}(2021)\citenamefont {Yan},
  \citenamefont {Xu}, \citenamefont {Kumar}, \citenamefont {Zhang},
  \citenamefont {Leng}, \citenamefont {Dunlap},\ and\ \citenamefont
  {Finzi}}]{yan21}%
  \BibitemOpen
  \bibfield  {author} {\bibinfo {author} {\bibfnamefont {Y.}~\bibnamefont
  {Yan}}, \bibinfo {author} {\bibfnamefont {W.}~\bibnamefont {Xu}}, \bibinfo
  {author} {\bibfnamefont {S.}~\bibnamefont {Kumar}}, \bibinfo {author}
  {\bibfnamefont {A.}~\bibnamefont {Zhang}}, \bibinfo {author} {\bibfnamefont
  {F.}~\bibnamefont {Leng}}, \bibinfo {author} {\bibfnamefont {D.}~\bibnamefont
  {Dunlap}},\ and\ \bibinfo {author} {\bibfnamefont {L.}~\bibnamefont
  {Finzi}},\ }\bibfield  {title} {\bibinfo {title} {Negative {DNA} supercoiling
  makes protein-mediated looping deterministic and ergodic within the bacterial
  doubling time},\ }\href {https://doi.org/10.1093/nar/gkab946} {\bibfield
  {journal} {\bibinfo  {journal} {Nucl. Acids Res.}\ }\textbf {\bibinfo
  {volume} {49}},\ \bibinfo {pages} {11550} (\bibinfo {year}
  {2021})}\BibitemShut {NoStop}%
\bibitem [{\citenamefont {Marko}\ and\ \citenamefont
  {Siggia}(1995{\natexlab{a}})}]{mark95b}%
  \BibitemOpen
  \bibfield  {author} {\bibinfo {author} {\bibfnamefont {J.~F.}\ \bibnamefont
  {Marko}}\ and\ \bibinfo {author} {\bibfnamefont {E.~D.}\ \bibnamefont
  {Siggia}},\ }\bibfield  {title} {\bibinfo {title} {{Statistical mechanics of
  supercoiled DNA}},\ }\href
  {https://doi.org/https://doi.org/10.1103/physreve.52.2912} {\bibfield
  {journal} {\bibinfo  {journal} {Phys. Rev. E}\ }\textbf {\bibinfo {volume}
  {52}},\ \bibinfo {pages} {2912} (\bibinfo {year}
  {1995}{\natexlab{a}})}\BibitemShut {NoStop}%
\bibitem [{\citenamefont {Marko}(2007)}]{mark07}%
  \BibitemOpen
  \bibfield  {author} {\bibinfo {author} {\bibfnamefont {J.~F.}\ \bibnamefont
  {Marko}},\ }\bibfield  {title} {\bibinfo {title} {{Torque and dynamics of
  linking number relaxation in stretched supercoiled DNA}},\ }\href
  {https://doi.org/https://doi.org/10.1103/PhysRevE.76.021926} {\bibfield
  {journal} {\bibinfo  {journal} {Phys. Rev. E}\ }\textbf {\bibinfo {volume}
  {76}},\ \bibinfo {pages} {021926} (\bibinfo {year} {2007})}\BibitemShut
  {NoStop}%
\bibitem [{\citenamefont {Vologodskii}\ \emph {et~al.}(1979)\citenamefont
  {Vologodskii}, \citenamefont {Anshelevich}, \citenamefont {Lukashin},\ and\
  \citenamefont {Frank-Kamenetskii}}]{volo79}%
  \BibitemOpen
  \bibfield  {author} {\bibinfo {author} {\bibfnamefont {A.~V.}\ \bibnamefont
  {Vologodskii}}, \bibinfo {author} {\bibfnamefont {V.~V.}\ \bibnamefont
  {Anshelevich}}, \bibinfo {author} {\bibfnamefont {A.~V.}\ \bibnamefont
  {Lukashin}},\ and\ \bibinfo {author} {\bibfnamefont {M.~D.}\ \bibnamefont
  {Frank-Kamenetskii}},\ }\bibfield  {title} {\bibinfo {title} {Statistical
  mechanics of supercoils and the torsional stiffness of the {DNA} double
  helix},\ }\href {https://doi.org/https://doi.org/10.1038/280294a0} {\bibfield
   {journal} {\bibinfo  {journal} {Nature}\ }\textbf {\bibinfo {volume}
  {280}},\ \bibinfo {pages} {294} (\bibinfo {year} {1979})}\BibitemShut
  {NoStop}%
\bibitem [{\citenamefont {Klenin}\ \emph {et~al.}(1991)\citenamefont {Klenin},
  \citenamefont {Vologodskii}, \citenamefont {Anshelevich}, \citenamefont
  {Dykhne},\ and\ \citenamefont {Frank-Kamenetskii}}]{klen91}%
  \BibitemOpen
  \bibfield  {author} {\bibinfo {author} {\bibfnamefont {K.~V.}\ \bibnamefont
  {Klenin}}, \bibinfo {author} {\bibfnamefont {A.~V.}\ \bibnamefont
  {Vologodskii}}, \bibinfo {author} {\bibfnamefont {V.~V.}\ \bibnamefont
  {Anshelevich}}, \bibinfo {author} {\bibfnamefont {A.~M.}\ \bibnamefont
  {Dykhne}},\ and\ \bibinfo {author} {\bibfnamefont {M.~D.}\ \bibnamefont
  {Frank-Kamenetskii}},\ }\bibfield  {title} {\bibinfo {title} {{Computer
  Simulation of DNA Supercoiling}},\ }\href
  {https://doi.org/https://doi.org/10.1016/0022-2836(91)90745-R} {\bibfield
  {journal} {\bibinfo  {journal} {J. Mol. Biol.}\ }\textbf {\bibinfo {volume}
  {217}},\ \bibinfo {pages} {413} (\bibinfo {year} {1991})}\BibitemShut
  {NoStop}%
\bibitem [{\citenamefont {Vologodskii}\ \emph {et~al.}(1992)\citenamefont
  {Vologodskii}, \citenamefont {Levene}, \citenamefont {Klenin}, \citenamefont
  {Frank-Kamenetskii},\ and\ \citenamefont {Cozzarelli}}]{volo92}%
  \BibitemOpen
  \bibfield  {author} {\bibinfo {author} {\bibfnamefont {A.~V.}\ \bibnamefont
  {Vologodskii}}, \bibinfo {author} {\bibfnamefont {S.~D.}\ \bibnamefont
  {Levene}}, \bibinfo {author} {\bibfnamefont {K.~V.}\ \bibnamefont {Klenin}},
  \bibinfo {author} {\bibfnamefont {M.}~\bibnamefont {Frank-Kamenetskii}},\
  and\ \bibinfo {author} {\bibfnamefont {N.~R.}\ \bibnamefont {Cozzarelli}},\
  }\bibfield  {title} {\bibinfo {title} {Conformational and thermodynamic
  properties of supercoiled {DNA}},\ }\href
  {https://doi.org/https://doi.org/10.1016/0022-2836(92)90533-P} {\bibfield
  {journal} {\bibinfo  {journal} {J. Mol. Biol.}\ }\textbf {\bibinfo {volume}
  {227}},\ \bibinfo {pages} {1224} (\bibinfo {year} {1992})}\BibitemShut
  {NoStop}%
\bibitem [{\citenamefont {Marko}\ and\ \citenamefont {Siggia}(1994)}]{mark94}%
  \BibitemOpen
  \bibfield  {author} {\bibinfo {author} {\bibfnamefont {J.~F.}\ \bibnamefont
  {Marko}}\ and\ \bibinfo {author} {\bibfnamefont {E.~D.}\ \bibnamefont
  {Siggia}},\ }\bibfield  {title} {\bibinfo {title} {Bending and twisting
  elasticity of {DNA}},\ }\href
  {https://doi.org/https://doi.org/10.1021/ma00082a015} {\bibfield  {journal}
  {\bibinfo  {journal} {Macromolecules}\ }\textbf {\bibinfo {volume} {27}},\
  \bibinfo {pages} {981} (\bibinfo {year} {1994})}\BibitemShut {NoStop}%
\bibitem [{\citenamefont {Strick}\ \emph {et~al.}(1996)\citenamefont {Strick},
  \citenamefont {Allemand}, \citenamefont {Bensimon}, \citenamefont
  {Bensimon},\ and\ \citenamefont {Croquette}}]{stri96}%
  \BibitemOpen
  \bibfield  {author} {\bibinfo {author} {\bibfnamefont {T.~R.}\ \bibnamefont
  {Strick}}, \bibinfo {author} {\bibfnamefont {J.~F.}\ \bibnamefont
  {Allemand}}, \bibinfo {author} {\bibfnamefont {D.}~\bibnamefont {Bensimon}},
  \bibinfo {author} {\bibfnamefont {A.}~\bibnamefont {Bensimon}},\ and\
  \bibinfo {author} {\bibfnamefont {V.}~\bibnamefont {Croquette}},\ }\bibfield
  {title} {\bibinfo {title} {The elasticity of a single supercoiled {DNA}
  molecule},\ }\href
  {https://doi.org/https://doi.org/10.1126/science.271.5257.1835} {\bibfield
  {journal} {\bibinfo  {journal} {Science}\ }\textbf {\bibinfo {volume}
  {271}},\ \bibinfo {pages} {1835} (\bibinfo {year} {1996})}\BibitemShut
  {NoStop}%
\bibitem [{\citenamefont {Forth}\ \emph {et~al.}(2008)\citenamefont {Forth},
  \citenamefont {Deufel}, \citenamefont {Sheinin}, \citenamefont {Daniels},
  \citenamefont {Sethna},\ and\ \citenamefont {Wang}}]{fort08}%
  \BibitemOpen
  \bibfield  {author} {\bibinfo {author} {\bibfnamefont {S.}~\bibnamefont
  {Forth}}, \bibinfo {author} {\bibfnamefont {C.}~\bibnamefont {Deufel}},
  \bibinfo {author} {\bibfnamefont {M.~Y.}\ \bibnamefont {Sheinin}}, \bibinfo
  {author} {\bibfnamefont {B.}~\bibnamefont {Daniels}}, \bibinfo {author}
  {\bibfnamefont {J.~P.}\ \bibnamefont {Sethna}},\ and\ \bibinfo {author}
  {\bibfnamefont {M.~D.}\ \bibnamefont {Wang}},\ }\bibfield  {title} {\bibinfo
  {title} {Abrupt buckling transition observed during the plectoneme formation
  of individual dna molecules},\ }\href
  {https://doi.org/https://doi.org/10.1103/PhysRevLett.100.148301} {\bibfield
  {journal} {\bibinfo  {journal} {Phys. Rev. Lett.}\ }\textbf {\bibinfo
  {volume} {100}},\ \bibinfo {pages} {148301} (\bibinfo {year}
  {2008})}\BibitemShut {NoStop}%
\bibitem [{\citenamefont {Wada}\ and\ \citenamefont {Netz}(2009)}]{wada09}%
  \BibitemOpen
  \bibfield  {author} {\bibinfo {author} {\bibfnamefont {H.}~\bibnamefont
  {Wada}}\ and\ \bibinfo {author} {\bibfnamefont {R.~R.}\ \bibnamefont
  {Netz}},\ }\bibfield  {title} {\bibinfo {title} {Plectoneme creation reduces
  the rotational friction of a polymer},\ }\href
  {https://doi.org/https://doi.org/10.1209/0295-5075/87/38001} {\bibfield
  {journal} {\bibinfo  {journal} {{EPL (Europhys. Lett.)}}\ }\textbf {\bibinfo
  {volume} {87}},\ \bibinfo {pages} {38001} (\bibinfo {year}
  {2009})}\BibitemShut {NoStop}%
\bibitem [{\citenamefont {Neukirch}\ and\ \citenamefont
  {Marko}(2011)}]{neuk11}%
  \BibitemOpen
  \bibfield  {author} {\bibinfo {author} {\bibfnamefont {S.}~\bibnamefont
  {Neukirch}}\ and\ \bibinfo {author} {\bibfnamefont {J.~F.}\ \bibnamefont
  {Marko}},\ }\bibfield  {title} {\bibinfo {title} {{Analytical description of
  extension, torque, and supercoiling radius of a stretched twisted DNA}},\
  }\href {https://doi.org/https://doi.org/10.1103/PhysRevLett.106.138104}
  {\bibfield  {journal} {\bibinfo  {journal} {Phys. Rev. Lett.}\ }\textbf
  {\bibinfo {volume} {106}},\ \bibinfo {pages} {138104} (\bibinfo {year}
  {2011})}\BibitemShut {NoStop}%
\bibitem [{\citenamefont {van Loenhout}\ \emph {et~al.}(2012)\citenamefont {van
  Loenhout}, \citenamefont {de~Grunt},\ and\ \citenamefont {Dekker}}]{vanl12}%
  \BibitemOpen
  \bibfield  {author} {\bibinfo {author} {\bibfnamefont {M.}~\bibnamefont {van
  Loenhout}}, \bibinfo {author} {\bibfnamefont {M.}~\bibnamefont {de~Grunt}},\
  and\ \bibinfo {author} {\bibfnamefont {C.}~\bibnamefont {Dekker}},\
  }\bibfield  {title} {\bibinfo {title} {Dynamics of {DNA} supercoils},\ }\href
  {https://doi.org/https://doi.org/10.1126/science.1225810} {\bibfield
  {journal} {\bibinfo  {journal} {Science}\ }\textbf {\bibinfo {volume}
  {338}},\ \bibinfo {pages} {94} (\bibinfo {year} {2012})}\BibitemShut
  {NoStop}%
\bibitem [{\citenamefont {Oberstrass}\ \emph {et~al.}(2012)\citenamefont
  {Oberstrass}, \citenamefont {Fernandes},\ and\ \citenamefont
  {Bryant}}]{ober12}%
  \BibitemOpen
  \bibfield  {author} {\bibinfo {author} {\bibfnamefont {F.~C.}\ \bibnamefont
  {Oberstrass}}, \bibinfo {author} {\bibfnamefont {L.~E.}\ \bibnamefont
  {Fernandes}},\ and\ \bibinfo {author} {\bibfnamefont {Z.}~\bibnamefont
  {Bryant}},\ }\bibfield  {title} {\bibinfo {title} {Torque measurements reveal
  sequence-specific cooperative transitions in supercoiled {DNA}},\ }\href
  {https://doi.org/https://doi.org/10.1073/pnas.1113532109} {\bibfield
  {journal} {\bibinfo  {journal} {Proc. Natl. Acad. Sci. USA}\ }\textbf
  {\bibinfo {volume} {109}},\ \bibinfo {pages} {6106} (\bibinfo {year}
  {2012})}\BibitemShut {NoStop}%
\bibitem [{\citenamefont {Lepage}\ \emph {et~al.}(2015)\citenamefont {Lepage},
  \citenamefont {K{\'{e}}p{\`{e}}s},\ and\ \citenamefont {Junier}}]{lepa15}%
  \BibitemOpen
  \bibfield  {author} {\bibinfo {author} {\bibfnamefont {T.}~\bibnamefont
  {Lepage}}, \bibinfo {author} {\bibfnamefont {F.}~\bibnamefont
  {K{\'{e}}p{\`{e}}s}},\ and\ \bibinfo {author} {\bibfnamefont
  {I.}~\bibnamefont {Junier}},\ }\bibfield  {title} {\bibinfo {title}
  {{Thermodynamics of Long Supercoiled Molecules: Insights from Highly
  Efficient {Monte Carlo} Simulations}},\ }\href
  {https://doi.org/10.1016/j.bpj.2015.06.005} {\bibfield  {journal} {\bibinfo
  {journal} {Biophys. J.}\ }\textbf {\bibinfo {volume} {109}},\ \bibinfo
  {pages} {135} (\bibinfo {year} {2015})}\BibitemShut {NoStop}%
\bibitem [{\citenamefont {Fathizadeh}\ \emph {et~al.}(2015)\citenamefont
  {Fathizadeh}, \citenamefont {Schiessel},\ and\ \citenamefont
  {Ejtehadi}}]{fath15}%
  \BibitemOpen
  \bibfield  {author} {\bibinfo {author} {\bibfnamefont {A.}~\bibnamefont
  {Fathizadeh}}, \bibinfo {author} {\bibfnamefont {H.}~\bibnamefont
  {Schiessel}},\ and\ \bibinfo {author} {\bibfnamefont {M.~R.}\ \bibnamefont
  {Ejtehadi}},\ }\bibfield  {title} {\bibinfo {title} {Molecular dynamics
  simulation of supercoiled {DNA} rings},\ }\href
  {https://doi.org/10.1021/ma501660w} {\bibfield  {journal} {\bibinfo
  {journal} {Macromolecules}\ }\textbf {\bibinfo {volume} {48}},\ \bibinfo
  {pages} {164} (\bibinfo {year} {2015})}\BibitemShut {NoStop}%
\bibitem [{\citenamefont {Benedetti}\ \emph {et~al.}(2015)\citenamefont
  {Benedetti}, \citenamefont {Japaridze}, \citenamefont {Dorier}, \citenamefont
  {Racko}, \citenamefont {Kwapich}, \citenamefont {Burnier}, \citenamefont
  {Dietler},\ and\ \citenamefont {Stasiak}}]{bene15}%
  \BibitemOpen
  \bibfield  {author} {\bibinfo {author} {\bibfnamefont {F.}~\bibnamefont
  {Benedetti}}, \bibinfo {author} {\bibfnamefont {A.}~\bibnamefont
  {Japaridze}}, \bibinfo {author} {\bibfnamefont {J.}~\bibnamefont {Dorier}},
  \bibinfo {author} {\bibfnamefont {D.}~\bibnamefont {Racko}}, \bibinfo
  {author} {\bibfnamefont {R.}~\bibnamefont {Kwapich}}, \bibinfo {author}
  {\bibfnamefont {Y.}~\bibnamefont {Burnier}}, \bibinfo {author} {\bibfnamefont
  {G.}~\bibnamefont {Dietler}},\ and\ \bibinfo {author} {\bibfnamefont
  {A.}~\bibnamefont {Stasiak}},\ }\bibfield  {title} {\bibinfo {title}
  {{Effects of physiological self-crowding of {DNA} on shape and biological
  properties of {DNA} molecules with various levels of supercoiling}},\ }\href
  {https://doi.org/10.1093/nar/gkv055} {\bibfield  {journal} {\bibinfo
  {journal} {Nucl. Acids Res.}\ }\textbf {\bibinfo {volume} {43}},\ \bibinfo
  {pages} {2390} (\bibinfo {year} {2015})}\BibitemShut {NoStop}%
\bibitem [{\citenamefont {Matek}\ \emph {et~al.}(2015)\citenamefont {Matek},
  \citenamefont {Ouldridge}, \citenamefont {Doye},\ and\ \citenamefont
  {Louis}}]{mate15}%
  \BibitemOpen
  \bibfield  {author} {\bibinfo {author} {\bibfnamefont {C.}~\bibnamefont
  {Matek}}, \bibinfo {author} {\bibfnamefont {T.~E.}\ \bibnamefont
  {Ouldridge}}, \bibinfo {author} {\bibfnamefont {J.~P.~K.}\ \bibnamefont
  {Doye}},\ and\ \bibinfo {author} {\bibfnamefont {A.~A.}\ \bibnamefont
  {Louis}},\ }\bibfield  {title} {\bibinfo {title} {Plectoneme tip bubbles:
  coupled denaturation and writhing in supercoiled {DNA}},\ }\href
  {https://doi.org/https://doi.org/10.1038/srep07655} {\bibfield  {journal}
  {\bibinfo  {journal} {Sci. Rep.}\ }\textbf {\bibinfo {volume} {5}},\ \bibinfo
  {pages} {7655} (\bibinfo {year} {2015})}\BibitemShut {NoStop}%
\bibitem [{\citenamefont {Ivenso}\ and\ \citenamefont
  {Lillian}(2016)}]{iven16}%
  \BibitemOpen
  \bibfield  {author} {\bibinfo {author} {\bibfnamefont {I.~D.}\ \bibnamefont
  {Ivenso}}\ and\ \bibinfo {author} {\bibfnamefont {T.~D.}\ \bibnamefont
  {Lillian}},\ }\bibfield  {title} {\bibinfo {title} {Simulation of {DNA}
  supercoil relaxation},\ }\href
  {https://doi.org/https://doi.org/10.1016/j.bpj.2016.03.041} {\bibfield
  {journal} {\bibinfo  {journal} {Biophys. J.}\ }\textbf {\bibinfo {volume}
  {110}},\ \bibinfo {pages} {2176} (\bibinfo {year} {2016})}\BibitemShut
  {NoStop}%
\bibitem [{\citenamefont {Barde}\ \emph {et~al.}(2018)\citenamefont {Barde},
  \citenamefont {Destainville},\ and\ \citenamefont {Manghi}}]{bard18}%
  \BibitemOpen
  \bibfield  {author} {\bibinfo {author} {\bibfnamefont {C.}~\bibnamefont
  {Barde}}, \bibinfo {author} {\bibfnamefont {N.}~\bibnamefont
  {Destainville}},\ and\ \bibinfo {author} {\bibfnamefont {M.}~\bibnamefont
  {Manghi}},\ }\bibfield  {title} {\bibinfo {title} {{Energy required to pinch
  a DNA plectoneme}},\ }\href
  {https://doi.org/https://doi.org/10.1103/PhysRevE.97.032412} {\bibfield
  {journal} {\bibinfo  {journal} {Phys. Rev. E}\ }\textbf {\bibinfo {volume}
  {97}},\ \bibinfo {pages} {032412} (\bibinfo {year} {2018})}\BibitemShut
  {NoStop}%
\bibitem [{\citenamefont {Fosado}\ \emph {et~al.}(2021)\citenamefont {Fosado},
  \citenamefont {Michieletto}, \citenamefont {Brackley},\ and\ \citenamefont
  {Marenduzzo}}]{fosa21}%
  \BibitemOpen
  \bibfield  {author} {\bibinfo {author} {\bibfnamefont {Y.~A.}\ \bibnamefont
  {Fosado}}, \bibinfo {author} {\bibfnamefont {D.}~\bibnamefont {Michieletto}},
  \bibinfo {author} {\bibfnamefont {C.~A.}\ \bibnamefont {Brackley}},\ and\
  \bibinfo {author} {\bibfnamefont {D.}~\bibnamefont {Marenduzzo}},\ }\bibfield
   {title} {\bibinfo {title} {{Nonequilibrium dynamics and action at a distance
  in transcriptionally driven DNA supercoiling}},\ }\href
  {https://doi.org/https://doi.org/10.1073/pnas.1905215118} {\bibfield
  {journal} {\bibinfo  {journal} {Proc. Natl. Acad. Sci. USA}\ }\textbf
  {\bibinfo {volume} {118}},\ \bibinfo {pages} {e1905215118} (\bibinfo {year}
  {2021})}\BibitemShut {NoStop}%
\bibitem [{\citenamefont {Ott}\ \emph {et~al.}(2020)\citenamefont {Ott},
  \citenamefont {Martini}, \citenamefont {Lipfert},\ and\ \citenamefont
  {Gerland}}]{ott20}%
  \BibitemOpen
  \bibfield  {author} {\bibinfo {author} {\bibfnamefont {K.}~\bibnamefont
  {Ott}}, \bibinfo {author} {\bibfnamefont {L.}~\bibnamefont {Martini}},
  \bibinfo {author} {\bibfnamefont {J.}~\bibnamefont {Lipfert}},\ and\ \bibinfo
  {author} {\bibfnamefont {U.}~\bibnamefont {Gerland}},\ }\bibfield  {title}
  {\bibinfo {title} {{Dynamics of the Buckling Transition in Double-Stranded
  DNA and RNA}},\ }\href {https://doi.org/10.1016/j.bpj.2020.01.049} {\bibfield
   {journal} {\bibinfo  {journal} {Biophys. J.}\ }\textbf {\bibinfo {volume}
  {118}},\ \bibinfo {pages} {1690} (\bibinfo {year} {2020})}\BibitemShut
  {NoStop}%
\bibitem [{\citenamefont {De~Vlaminck}\ and\ \citenamefont
  {Dekker}(2012)}]{devl12}%
  \BibitemOpen
  \bibfield  {author} {\bibinfo {author} {\bibfnamefont {I.}~\bibnamefont
  {De~Vlaminck}}\ and\ \bibinfo {author} {\bibfnamefont {C.}~\bibnamefont
  {Dekker}},\ }\bibfield  {title} {\bibinfo {title} {Recent advances in
  magnetic tweezers},\ }\href
  {https://doi.org/https://doi.org/10.1146/annurev-biophys-122311-100544}
  {\bibfield  {journal} {\bibinfo  {journal} {Annu. Rev. Biophys.}\ }\textbf
  {\bibinfo {volume} {41}},\ \bibinfo {pages} {453} (\bibinfo {year}
  {2012})}\BibitemShut {NoStop}%
\bibitem [{\citenamefont {Lipfert}\ \emph {et~al.}(2011)\citenamefont
  {Lipfert}, \citenamefont {Wiggin}, \citenamefont {Kerssemakers},
  \citenamefont {Pedaci},\ and\ \citenamefont {Dekker}}]{lipf11}%
  \BibitemOpen
  \bibfield  {author} {\bibinfo {author} {\bibfnamefont {J.}~\bibnamefont
  {Lipfert}}, \bibinfo {author} {\bibfnamefont {M.}~\bibnamefont {Wiggin}},
  \bibinfo {author} {\bibfnamefont {J.~W.~J.}\ \bibnamefont {Kerssemakers}},
  \bibinfo {author} {\bibfnamefont {F.}~\bibnamefont {Pedaci}},\ and\ \bibinfo
  {author} {\bibfnamefont {N.~H.}\ \bibnamefont {Dekker}},\ }\bibfield  {title}
  {\bibinfo {title} {Freely orbiting magnetic tweezers to directly monitor
  changes in the twist of nucleic acids},\ }\href
  {https://doi.org/https://doi.org/10.1038/ncomms1450} {\bibfield  {journal}
  {\bibinfo  {journal} {Nature Comm.}\ }\textbf {\bibinfo {volume} {2}},\
  \bibinfo {pages} {439} (\bibinfo {year} {2011})}\BibitemShut {NoStop}%
\bibitem [{\citenamefont {Vanderlinden}\ \emph {et~al.}(2021)\citenamefont
  {Vanderlinden}, \citenamefont {Skoruppa}, \citenamefont {Kolbeck},
  \citenamefont {Carlon},\ and\ \citenamefont {Lipfert}}]{vand21}%
  \BibitemOpen
  \bibfield  {author} {\bibinfo {author} {\bibfnamefont {W.}~\bibnamefont
  {Vanderlinden}}, \bibinfo {author} {\bibfnamefont {E.}~\bibnamefont
  {Skoruppa}}, \bibinfo {author} {\bibfnamefont {P.}~\bibnamefont {Kolbeck}},
  \bibinfo {author} {\bibfnamefont {E.}~\bibnamefont {Carlon}},\ and\ \bibinfo
  {author} {\bibfnamefont {J.}~\bibnamefont {Lipfert}},\ }\bibfield  {title}
  {\bibinfo {title} {{DNA fluctuations reveal the size and dynamics of
  topological domains}},\ }\bibfield  {journal} {\bibinfo  {journal} {BiorXiv
  preprint}\ }\href {https://doi.org/https://doi.org/10.1101/2021.12.21.473646}
  {https://doi.org/10.1101/2021.12.21.473646} (\bibinfo {year}
  {2021})\BibitemShut {NoStop}%
\bibitem [{\citenamefont {Marko}\ and\ \citenamefont
  {Neukirch}(2012)}]{mark12}%
  \BibitemOpen
  \bibfield  {author} {\bibinfo {author} {\bibfnamefont {J.~F.}\ \bibnamefont
  {Marko}}\ and\ \bibinfo {author} {\bibfnamefont {S.}~\bibnamefont
  {Neukirch}},\ }\bibfield  {title} {\bibinfo {title} {{Competition between
  curls and plectonemes near the buckling transition of stretched supercoiled
  DNA}},\ }\href {https://doi.org/10.1103/physreve.85.011908} {\bibfield
  {journal} {\bibinfo  {journal} {Phys. Rev. E}\ }\textbf {\bibinfo {volume}
  {85}},\ \bibinfo {pages} {011908} (\bibinfo {year} {2012})}\BibitemShut
  {NoStop}%
\bibitem [{\citenamefont {Emanuel}\ \emph {et~al.}(2013)\citenamefont
  {Emanuel}, \citenamefont {Lanzani},\ and\ \citenamefont
  {Schiessel}}]{eman13}%
  \BibitemOpen
  \bibfield  {author} {\bibinfo {author} {\bibfnamefont {M.}~\bibnamefont
  {Emanuel}}, \bibinfo {author} {\bibfnamefont {G.}~\bibnamefont {Lanzani}},\
  and\ \bibinfo {author} {\bibfnamefont {H.}~\bibnamefont {Schiessel}},\
  }\bibfield  {title} {\bibinfo {title} {Multiplectoneme phase of
  double-stranded {DNA} under torsion},\ }\href
  {https://doi.org/https://doi.org/10.1103/physreve.88.022706} {\bibfield
  {journal} {\bibinfo  {journal} {Phys. Rev. E}\ }\textbf {\bibinfo {volume}
  {88}},\ \bibinfo {pages} {022706} (\bibinfo {year} {2013})}\BibitemShut
  {NoStop}%
\bibitem [{\citenamefont {Moroz}\ and\ \citenamefont {Nelson}(1997)}]{moro97}%
  \BibitemOpen
  \bibfield  {author} {\bibinfo {author} {\bibfnamefont {J.~D.}\ \bibnamefont
  {Moroz}}\ and\ \bibinfo {author} {\bibfnamefont {P.}~\bibnamefont {Nelson}},\
  }\bibfield  {title} {\bibinfo {title} {Torsional directed walks, entropic
  elasticity, and {DNA} twist stiffness},\ }\href
  {https://doi.org/https://doi.org/10.1073/pnas.94.26.14418} {\bibfield
  {journal} {\bibinfo  {journal} {Proc. Natl. Acad. Sci. USA}\ }\textbf
  {\bibinfo {volume} {94}},\ \bibinfo {pages} {14418} (\bibinfo {year}
  {1997})}\BibitemShut {NoStop}%
\bibitem [{\citenamefont {Marko}\ and\ \citenamefont
  {Siggia}(1995{\natexlab{b}})}]{mark95}%
  \BibitemOpen
  \bibfield  {author} {\bibinfo {author} {\bibfnamefont {J.~F.}\ \bibnamefont
  {Marko}}\ and\ \bibinfo {author} {\bibfnamefont {E.~D.}\ \bibnamefont
  {Siggia}},\ }\bibfield  {title} {\bibinfo {title} {{Stretching DNA}},\
  }\href@noop {} {\bibfield  {journal} {\bibinfo  {journal} {Macromolecules}\
  }\textbf {\bibinfo {volume} {28}},\ \bibinfo {pages} {8759} (\bibinfo {year}
  {1995}{\natexlab{b}})}\BibitemShut {NoStop}%
\bibitem [{\citenamefont {Moroz}\ and\ \citenamefont {Nelson}(1998)}]{moro98}%
  \BibitemOpen
  \bibfield  {author} {\bibinfo {author} {\bibfnamefont {J.~D.}\ \bibnamefont
  {Moroz}}\ and\ \bibinfo {author} {\bibfnamefont {P.}~\bibnamefont {Nelson}},\
  }\bibfield  {title} {\bibinfo {title} {Entropic elasticity of twist-storing
  polymers},\ }\href {https://doi.org/https://doi.org/10.1021/ma971804a}
  {\bibfield  {journal} {\bibinfo  {journal} {Macromolecules}\ }\textbf
  {\bibinfo {volume} {31}},\ \bibinfo {pages} {6333} (\bibinfo {year}
  {1998})}\BibitemShut {NoStop}%
\bibitem [{\citenamefont {Skoruppa}\ \emph {et~al.}(2017)\citenamefont
  {Skoruppa}, \citenamefont {Laleman}, \citenamefont {Nomidis},\ and\
  \citenamefont {Carlon}}]{skor17}%
  \BibitemOpen
  \bibfield  {author} {\bibinfo {author} {\bibfnamefont {E.}~\bibnamefont
  {Skoruppa}}, \bibinfo {author} {\bibfnamefont {M.}~\bibnamefont {Laleman}},
  \bibinfo {author} {\bibfnamefont {S.~K.}\ \bibnamefont {Nomidis}},\ and\
  \bibinfo {author} {\bibfnamefont {E.}~\bibnamefont {Carlon}},\ }\bibfield
  {title} {\bibinfo {title} {{DNA elasticity from coarse-grained simulations:
  The effect of groove asymmetry}},\ }\href
  {https://doi.org/http://aip.scitation.org/doi/10.1063/1.4984039} {\bibfield
  {journal} {\bibinfo  {journal} {J. Chem. Phys.}\ }\textbf {\bibinfo {volume}
  {146}},\ \bibinfo {pages} {214902} (\bibinfo {year} {2017})}\BibitemShut
  {NoStop}%
\bibitem [{\citenamefont {Skoruppa}\ \emph {et~al.}(2018)\citenamefont
  {Skoruppa}, \citenamefont {Nomidis}, \citenamefont {Marko},\ and\
  \citenamefont {Carlon}}]{skor18}%
  \BibitemOpen
  \bibfield  {author} {\bibinfo {author} {\bibfnamefont {E.}~\bibnamefont
  {Skoruppa}}, \bibinfo {author} {\bibfnamefont {S.~K.}\ \bibnamefont
  {Nomidis}}, \bibinfo {author} {\bibfnamefont {J.~F.}\ \bibnamefont {Marko}},\
  and\ \bibinfo {author} {\bibfnamefont {E.}~\bibnamefont {Carlon}},\
  }\bibfield  {title} {\bibinfo {title} {{Bend-Induced Twist Waves and the
  Structure of Nucleosomal DNA}},\ }\href
  {https://doi.org/https://doi.org/10.1103/PhysRevLett.121.088101} {\bibfield
  {journal} {\bibinfo  {journal} {Phys. Rev. Lett.}\ }\textbf {\bibinfo
  {volume} {121}},\ \bibinfo {pages} {088101} (\bibinfo {year}
  {2018})}\BibitemShut {NoStop}%
\bibitem [{\citenamefont {Skoruppa}\ \emph {et~al.}(2021)\citenamefont
  {Skoruppa}, \citenamefont {Voorspoels}, \citenamefont {Vreede},\ and\
  \citenamefont {Carlon}}]{skor21}%
  \BibitemOpen
  \bibfield  {author} {\bibinfo {author} {\bibfnamefont {E.}~\bibnamefont
  {Skoruppa}}, \bibinfo {author} {\bibfnamefont {A.}~\bibnamefont
  {Voorspoels}}, \bibinfo {author} {\bibfnamefont {J.}~\bibnamefont {Vreede}},\
  and\ \bibinfo {author} {\bibfnamefont {E.}~\bibnamefont {Carlon}},\
  }\bibfield  {title} {\bibinfo {title} {Length-scale-dependent elasticity in
  {DNA} from coarse-grained and all-atom models},\ }\href
  {https://doi.org/https://doi.org/10.1103/physreve.103.042408} {\bibfield
  {journal} {\bibinfo  {journal} {Phys. Rev. E}\ }\textbf {\bibinfo {volume}
  {103}},\ \bibinfo {pages} {042408} (\bibinfo {year} {2021})}\BibitemShut
  {NoStop}%
\bibitem [{\citenamefont {Bouchiat}\ and\ \citenamefont
  {M\'ezard}(1998)}]{bouc98}%
  \BibitemOpen
  \bibfield  {author} {\bibinfo {author} {\bibfnamefont {C.}~\bibnamefont
  {Bouchiat}}\ and\ \bibinfo {author} {\bibfnamefont {M.}~\bibnamefont
  {M\'ezard}},\ }\bibfield  {title} {\bibinfo {title} {Elasticity model of a
  supercoiled {DNA} molecule},\ }\href
  {https://doi.org/https://doi.org/10.1103/PhysRevLett.80.1556} {\bibfield
  {journal} {\bibinfo  {journal} {Phys. Rev. Lett.}\ }\textbf {\bibinfo
  {volume} {80}},\ \bibinfo {pages} {1556} (\bibinfo {year}
  {1998})}\BibitemShut {NoStop}%
\bibitem [{\citenamefont {Bouchiat}\ and\ \citenamefont
  {M{\'{e}}zard}(2000)}]{bouc00}%
  \BibitemOpen
  \bibfield  {author} {\bibinfo {author} {\bibfnamefont {C.}~\bibnamefont
  {Bouchiat}}\ and\ \bibinfo {author} {\bibfnamefont {M.}~\bibnamefont
  {M{\'{e}}zard}},\ }\bibfield  {title} {\bibinfo {title} {Elastic rod model of
  a supercoiled {DNA} molecule},\ }\href
  {https://doi.org/10.1007/s101890050020} {\bibfield  {journal} {\bibinfo
  {journal} {Eur. Phys. J. E}\ }\textbf {\bibinfo {volume} {2}},\ \bibinfo
  {pages} {377} (\bibinfo {year} {2000})}\BibitemShut {NoStop}%
\bibitem [{\citenamefont {Gao}\ \emph {et~al.}(2021)\citenamefont {Gao},
  \citenamefont {Hong}, \citenamefont {Ye}, \citenamefont {Inman},\ and\
  \citenamefont {Wang}}]{gao21}%
  \BibitemOpen
  \bibfield  {author} {\bibinfo {author} {\bibfnamefont {X.}~\bibnamefont
  {Gao}}, \bibinfo {author} {\bibfnamefont {Y.}~\bibnamefont {Hong}}, \bibinfo
  {author} {\bibfnamefont {F.}~\bibnamefont {Ye}}, \bibinfo {author}
  {\bibfnamefont {J.~T.}\ \bibnamefont {Inman}},\ and\ \bibinfo {author}
  {\bibfnamefont {M.~D.}\ \bibnamefont {Wang}},\ }\bibfield  {title} {\bibinfo
  {title} {Torsional stiffness of extended and plectonemic dna},\ }\href
  {https://doi.org/10.1103/PhysRevLett.127.028101} {\bibfield  {journal}
  {\bibinfo  {journal} {Phys. Rev. Lett.}\ }\textbf {\bibinfo {volume} {127}},\
  \bibinfo {pages} {028101} (\bibinfo {year} {2021})}\BibitemShut {NoStop}%
\bibitem [{\citenamefont {Rybenkov}\ \emph {et~al.}(1993)\citenamefont
  {Rybenkov}, \citenamefont {Cozzarelli},\ and\ \citenamefont
  {Vologodskii}}]{rybe93}%
  \BibitemOpen
  \bibfield  {author} {\bibinfo {author} {\bibfnamefont {V.~V.}\ \bibnamefont
  {Rybenkov}}, \bibinfo {author} {\bibfnamefont {N.~R.}\ \bibnamefont
  {Cozzarelli}},\ and\ \bibinfo {author} {\bibfnamefont {A.~V.}\ \bibnamefont
  {Vologodskii}},\ }\bibfield  {title} {\bibinfo {title} {Probability of {DNA}
  knotting and the effective diameter of the {DNA} double helix},\ }\href
  {https://doi.org/https://doi.org/10.1073/pnas.90.11.5307} {\bibfield
  {journal} {\bibinfo  {journal} {Proc. Natl. Acad. Sci. USA}\ }\textbf
  {\bibinfo {volume} {90}},\ \bibinfo {pages} {5307} (\bibinfo {year}
  {1993})}\BibitemShut {NoStop}%
\bibitem [{\citenamefont {Frenkel}\ and\ \citenamefont {Smit}(2002)}]{fren02}%
  \BibitemOpen
  \bibfield  {author} {\bibinfo {author} {\bibfnamefont {D.}~\bibnamefont
  {Frenkel}}\ and\ \bibinfo {author} {\bibfnamefont {B.}~\bibnamefont {Smit}},\
  }\href@noop {} {\emph {\bibinfo {title} {Understanding Molecular Simulation:
  From Algorithms to Applications}}},\ \bibinfo {edition} {2nd}\ ed.,\ \bibinfo
  {series} {Computational Science Series}, Vol.~\bibinfo {volume} {1}\
  (\bibinfo  {publisher} {Academic Press},\ \bibinfo {address} {San Diego},\
  \bibinfo {year} {2002})\BibitemShut {NoStop}%
\bibitem [{\citenamefont {Kumar}\ \emph {et~al.}(1992)\citenamefont {Kumar},
  \citenamefont {Rosenberg}, \citenamefont {Bouzida}, \citenamefont
  {Swendsen},\ and\ \citenamefont {Kollman}}]{kuma92}%
  \BibitemOpen
  \bibfield  {author} {\bibinfo {author} {\bibfnamefont {S.}~\bibnamefont
  {Kumar}}, \bibinfo {author} {\bibfnamefont {J.~M.}\ \bibnamefont
  {Rosenberg}}, \bibinfo {author} {\bibfnamefont {D.}~\bibnamefont {Bouzida}},
  \bibinfo {author} {\bibfnamefont {R.~H.}\ \bibnamefont {Swendsen}},\ and\
  \bibinfo {author} {\bibfnamefont {P.~A.}\ \bibnamefont {Kollman}},\
  }\bibfield  {title} {\bibinfo {title} {The weighted histogram analysis method
  for free-energy calculations on biomolecules. i. the method},\ }\href
  {https://doi.org/10.1002/jcc.540130812} {\bibfield  {journal} {\bibinfo
  {journal} {J. Comp. Chem.}\ }\textbf {\bibinfo {volume} {13}},\ \bibinfo
  {pages} {1011} (\bibinfo {year} {1992})}\BibitemShut {NoStop}%
\bibitem [{\citenamefont {Liu}\ and\ \citenamefont {Chan}(2008)}]{liu08a}%
  \BibitemOpen
  \bibfield  {author} {\bibinfo {author} {\bibfnamefont {Z.}~\bibnamefont
  {Liu}}\ and\ \bibinfo {author} {\bibfnamefont {H.~S.}\ \bibnamefont {Chan}},\
  }\bibfield  {title} {\bibinfo {title} {{Efficient chain moves for Monte Carlo
  simulations of a wormlike DNA model: Excluded volume, supercoils, site
  juxtapositions, knots, and comparisons with random-flight and lattice
  models}},\ }\href {https://doi.org/10.1063/1.2899022} {\bibfield  {journal}
  {\bibinfo  {journal} {J. Chem. Phys.}\ }\textbf {\bibinfo {volume} {128}},\
  \bibinfo {pages} {145104} (\bibinfo {year} {2008})}\BibitemShut {NoStop}%
\bibitem [{\citenamefont {Coronel}\ \emph {et~al.}(2018)\citenamefont
  {Coronel}, \citenamefont {Suma},\ and\ \citenamefont {Micheletti}}]{coro18}%
  \BibitemOpen
  \bibfield  {author} {\bibinfo {author} {\bibfnamefont {L.}~\bibnamefont
  {Coronel}}, \bibinfo {author} {\bibfnamefont {A.}~\bibnamefont {Suma}},\ and\
  \bibinfo {author} {\bibfnamefont {C.}~\bibnamefont {Micheletti}},\ }\bibfield
   {title} {\bibinfo {title} {{Dynamics of supercoiled DNA with complex knots:
  large-scale rearrangements and persistent multi-strand interlocking}},\
  }\href {https://doi.org/10.1093/nar/gky523} {\bibfield  {journal} {\bibinfo
  {journal} {Nucl. Acids Res.}\ }\textbf {\bibinfo {volume} {46}},\ \bibinfo
  {pages} {7533} (\bibinfo {year} {2018})}\BibitemShut {NoStop}%
\bibitem [{\citenamefont {Desai}\ \emph {et~al.}(2020)\citenamefont {Desai},
  \citenamefont {Brahmachari}, \citenamefont {Marko}, \citenamefont {Das},\
  and\ \citenamefont {Neuman}}]{desa20}%
  \BibitemOpen
  \bibfield  {author} {\bibinfo {author} {\bibfnamefont {P.~R.}\ \bibnamefont
  {Desai}}, \bibinfo {author} {\bibfnamefont {S.}~\bibnamefont {Brahmachari}},
  \bibinfo {author} {\bibfnamefont {J.~F.}\ \bibnamefont {Marko}}, \bibinfo
  {author} {\bibfnamefont {S.}~\bibnamefont {Das}},\ and\ \bibinfo {author}
  {\bibfnamefont {K.~C.}\ \bibnamefont {Neuman}},\ }\bibfield  {title}
  {\bibinfo {title} {{Coarse-grained modelling of DNA plectoneme pinning in the
  presence of base-pair mismatches}},\ }\href
  {https://doi.org/10.1093/nar/gkaa836} {\bibfield  {journal} {\bibinfo
  {journal} {Nucl. Acids Res.}\ }\textbf {\bibinfo {volume} {48}},\ \bibinfo
  {pages} {10713} (\bibinfo {year} {2020})}\BibitemShut {NoStop}%
\bibitem [{\citenamefont {Fuller}(1971)}]{full71}%
  \BibitemOpen
  \bibfield  {author} {\bibinfo {author} {\bibfnamefont {F.~B.}\ \bibnamefont
  {Fuller}},\ }\bibfield  {title} {\bibinfo {title} {{The Writhing Number of a
  Space Curve}},\ }\href
  {https://doi.org/https://doi.org/10.1073/pnas.68.4.815} {\bibfield  {journal}
  {\bibinfo  {journal} {Proc. Natl. Acad. Sci. USA}\ }\textbf {\bibinfo
  {volume} {68}},\ \bibinfo {pages} {815} (\bibinfo {year} {1971})}\BibitemShut
  {NoStop}%
\bibitem [{\citenamefont {Marko}(2015)}]{mark15}%
  \BibitemOpen
  \bibfield  {author} {\bibinfo {author} {\bibfnamefont {J.~F.}\ \bibnamefont
  {Marko}},\ }\bibfield  {title} {\bibinfo {title} {Biophysics of protein-{DNA}
  interactions and chromosome organization},\ }\href
  {https://doi.org/https://doi.org/10.1016/j.physa.2014.07.045} {\bibfield
  {journal} {\bibinfo  {journal} {Physica A}\ }\textbf {\bibinfo {volume}
  {418}},\ \bibinfo {pages} {126} (\bibinfo {year} {2015})}\BibitemShut
  {NoStop}%
\bibitem [{Note1()}]{Note1}%
  \BibitemOpen
  \bibinfo {note} {We remark that writhe is technically only defined for a
  closed space curve, which in the context of stretched linear DNA is usually
  resolved by introducing a virtual closure at either infinity or through a
  large arc \cite {volo97,chou14}. We ignore these closure contributions, since
  they are irrelevant for the detection of plectonemes.}\BibitemShut {Stop}%
\bibitem [{\citenamefont {Klenin}\ and\ \citenamefont
  {Langowski}(2000)}]{klen00}%
  \BibitemOpen
  \bibfield  {author} {\bibinfo {author} {\bibfnamefont {K.}~\bibnamefont
  {Klenin}}\ and\ \bibinfo {author} {\bibfnamefont {J.}~\bibnamefont
  {Langowski}},\ }\bibfield  {title} {\bibinfo {title} {{Computation of writhe
  in modeling of supercoiled DNA}},\ }\href
  {https://doi.org/10.1002/1097-0282(20001015)54:5<307::AID-BIP20>3.0.CO;2-Y}
  {\bibfield  {journal} {\bibinfo  {journal} {Biopolymers}\ }\textbf {\bibinfo
  {volume} {54}},\ \bibinfo {pages} {307} (\bibinfo {year} {2000})}\BibitemShut
  {NoStop}%
\bibitem [{\citenamefont {Nomidis}(2020)}]{nomi20_thesis}%
  \BibitemOpen
  \bibfield  {author} {\bibinfo {author} {\bibfnamefont {S.~K.}\ \bibnamefont
  {Nomidis}},\ }\emph {\bibinfo {title} {Theory and simulation of DNA mechanics
  and hybridization}},\ \href@noop {} {Ph.D. thesis},\ \bibinfo  {school} {KU
  Leuven} (\bibinfo {year} {2020})\BibitemShut {NoStop}%
\bibitem [{\citenamefont {Yang}\ \emph {et~al.}(2000)\citenamefont {Yang},
  \citenamefont {Haijun},\ and\ \citenamefont {Zhong-Can}}]{yang00}%
  \BibitemOpen
  \bibfield  {author} {\bibinfo {author} {\bibfnamefont {Z.}~\bibnamefont
  {Yang}}, \bibinfo {author} {\bibfnamefont {Z.}~\bibnamefont {Haijun}},\ and\
  \bibinfo {author} {\bibfnamefont {O.~Y.}\ \bibnamefont {Zhong-Can}},\
  }\bibfield  {title} {\bibinfo {title} {{Monte Carlo implementation of
  supercoiled double-stranded DNA}},\ }\href
  {https://doi.org/10.1016/S0006-3495(00)76745-2} {\bibfield  {journal}
  {\bibinfo  {journal} {Biophys. J.}\ }\textbf {\bibinfo {volume} {78}},\
  \bibinfo {pages} {1979} (\bibinfo {year} {2000})}\BibitemShut {NoStop}%
\bibitem [{\citenamefont {Vologodskii}\ and\ \citenamefont
  {Marko}(1997)}]{volo97}%
  \BibitemOpen
  \bibfield  {author} {\bibinfo {author} {\bibfnamefont {A.~V.}\ \bibnamefont
  {Vologodskii}}\ and\ \bibinfo {author} {\bibfnamefont {J.~F.}\ \bibnamefont
  {Marko}},\ }\bibfield  {title} {\bibinfo {title} {Extension of torsionally
  stressed {DNA} by external force},\ }\href
  {https://doi.org/https://doi.org/10.1016/S0006-3495(97)78053-6} {\bibfield
  {journal} {\bibinfo  {journal} {Biophys. J.}\ }\textbf {\bibinfo {volume}
  {73}},\ \bibinfo {pages} {123} (\bibinfo {year} {1997})}\BibitemShut
  {NoStop}%
\bibitem [{\citenamefont {Chou}\ \emph {et~al.}(2014)\citenamefont {Chou},
  \citenamefont {Lipfert},\ and\ \citenamefont {Das}}]{chou14}%
  \BibitemOpen
  \bibfield  {author} {\bibinfo {author} {\bibfnamefont {F.~C.}\ \bibnamefont
  {Chou}}, \bibinfo {author} {\bibfnamefont {J.}~\bibnamefont {Lipfert}},\ and\
  \bibinfo {author} {\bibfnamefont {R.}~\bibnamefont {Das}},\ }\bibfield
  {title} {\bibinfo {title} {{Blind predictions of DNA and RNA tweezers
  experiments with force and torque}},\ }\href
  {https://doi.org/https://doi.org/10.1371/journal.pcbi.1003756} {\bibfield
  {journal} {\bibinfo  {journal} {PLoS Comp. Biol.}\ }\textbf {\bibinfo
  {volume} {10}},\ \bibinfo {pages} {e1003756} (\bibinfo {year}
  {2014})}\BibitemShut {NoStop}%
\end{thebibliography}%

\end{document}